\documentclass[12pt]{article}

\usepackage{rotating}
\usepackage{axodraw}
\usepackage{latexsym}
\def\nn{\nonumber}
\def\a{& \hspace{-7pt}}

\def\Z{{\bf Z}}

\def\ds{\displaystyle}

\def\R{\mathcal R}
\def\mc{\mathcal}
\def\ZZ{Z\hspace{-7pt}Z}
\def\b{\raisebox{13pt}{}\raisebox{-6pt}{}$\!\!$}
\def\d{\raisebox{12pt}{}\raisebox{-5pt}{}$\!\!$}
\def\tl{\tilde}

\renewcommand{\section}[1]{\setcounter{equation}{0}
 \addtocounter{section}{1}
 \vspace{5mm} \par \noindent {\large \bf \thesection . #1}
 \setcounter{subsection}{0} \par \vspace{2mm} }

\renewcommand{\subsection}[1]{\addtocounter{subsection}{1}
 \vspace{2.5mm}\par\noindent {\bf  \thesubsection . #1}\par
 \vspace{0.5mm}}
 \renewcommand{\thebibliography}[1]{{\vspace{5mm}\par
  \noindent{\large \bf References}\par \vspace{2mm}}
  \list{$[$\arabic{enumi}$]$}{\settowidth\labelwidth{[#1]}
  \leftmargin \labelwidth \advance\leftmargin\labelsep
  \addtolength{\topsep}{-4em}\usecounter{enumi}}
  \def\newblock{\hskip .11em plus .33em minus .07em}
  \sloppy\clubpenalty4000\widowpenalty4000 \sfcode`\.=1000\relax
  \setlength{\itemsep}{-0.4em}}
\newcommand{\acknowledgments}[1]{\vspace{5mm}\par
 \noindent{\large \bf Acknowledgments}\par \vspace{2mm}}

\begin{document}

\thispagestyle{empty}

\begin{center}
\hfill CERN-TH/2003-315\\
\hfill ROMA-1365/03 \\
\hfill SISSA-107/2003/EP \\

\begin{center}

\vspace{1.7cm}

{\LARGE\bf Gauge-Higgs Unification \\[3mm]
in Orbifold Models}

\end{center}

\vspace{1.4cm}

{\bf C. A. Scrucca$^{a}$, M. Serone$^{b}$,
L. Silvestrini$^{c}$, A. Wulzer$^{b}$}\\

\vspace{1.2cm}

${}^a\!\!$
{\em Theor. Phys. Div., CERN, CH-1211 Geneva 23, Switzerland}
\vspace{.3cm}

${}^b\!\!$
{\em ISAS-SISSA and INFN, Via Beirut 2-4, I-34013 Trieste, Italy}
\vspace{.3cm}

${}^c\!\!$
{\em INFN, Sez. di Roma, Dip. di Fisica, Univ. di Roma ``La Sapienza''} \\
{\em P.le Aldo Moro 2, I-00185, Rome, Italy}
\end{center}

\vspace{0.8cm}

\centerline{\bf Abstract}
\vspace{2 mm}
\begin{quote}\small

Six-dimensional orbifold models where the Higgs field is identified
with some internal component of a gauge field are considered.
We classify all possible $T^2/\Z_N$ orbifold constructions based
on a $SU(3)$ electroweak gauge symmetry. Depending on the orbifold
twist, models with two, one or zero Higgs doublets can be obtained.
Models with one Higgs doublet are particularly interesting, as they
lead to a prediction for the Higgs mass that is twice the $W$ boson
mass at leading order: $m_H=2\,m_W$.
The electroweak scale is quadratically sensitive to
the cut-off, but only through very specific localized operators.
We study in detail the structure of these operators at one loop,
and identify a class of models where they do not destabilize the
electroweak scale at the leading order. This provides a very promising
framework to construct realistic and predictive models of electroweak
symmetry breaking.

\end{quote}

\vfill

\newpage

\section{Introduction}

If the Standard Model (SM) is seen as a low-energy effective
description of a more fundamental theory valid up to a scale
$\Lambda$, the order of magnitude of the Higgs mass $m_H$, as
suggested by global fits of electroweak precision data, is
natural only if one assumes a quite low scale for $\Lambda$.
Taking as a reference value $m_H \sim$ 100 GeV and requiring
this to be the magnitude of the leading one-loop correction
from the Yukawa coupling to the top quark,
$\delta m_H \sim \frac {\sqrt{3}}{2\pi} |\lambda_t| \Lambda$,
one finds $\Lambda \sim 400$ GeV. On the other hand, in order
to respect the stringent bounds from electroweak precision
physics, $\Lambda$ should be much higher than that.
Present bounds from generic four-fermion operators with
coefficients of order $\Lambda^{-2}$ require $\Lambda$ to be at
least $5-10$ TeV (see {\em e.g}.~\cite{Barbieri:1999tm}).
This discrepancy of more than
one order of magnitude between the natural value of $\Lambda$
and its experimental lower bound defines the little hierarchy
problem, and can be interpreted as a measure of the amount of
fine-tuning that is required for a generic extension of the SM.
Its present value of about $5-10\,\%$ poses a significant
theoretical problem, which is known to affect also the most
promising scenario of physics beyond the SM, namely Supersymmetry
(SUSY). This strongly motivates the investigation of other possible
scenarios which could solve this little hierarchy problem.

The idea that the SM Higgs field might be an internal component
of a gauge field of an extended electroweak symmetry, propagating
in more than four dimensions, is particularly appealing in the above
context, because it allows to build models where the electroweak
symmetry breaking (EWSB) scale is stabilized thanks to the
higher-dimensional gauge symmetry. This idea of gauge--Higgs
unification was proposed long ago \cite{early}, and recently
received renewed interest, in both its non-SUSY
\cite{Hatanaka:1998yp}--\cite{Scrucca:2003ra} and SUSY
\cite{Hall:2001zb} versions.
The simplest framework allowing its implementation is a five-dimensional
($5D$) $SU(3)$ gauge theory on an $S^1/\Z_2$ orbifold \cite{Antoniadis:1993jp}.
This model presents many interesting features, but it predicts, in
its minimal version, too low values for the Higgs mass (see
\cite{Scrucca:2003ra} for a detailed study of these models), because
of the absence of any tree-level Higgs potential, a common feature
of all $5D$ models with a single Higgs doublet.\footnote{In $5D$ SUSY models,
a tree-level Higgs potential can occur, but at the price of having at
least two Higgs doublets.}

New features emerge when applying the above ideas in the presence of
two or more extra dimensions.\footnote{$D>5$ orbifolds also present
new possibilities in the context of flavour physics, see
\textit{e.g.} ref.~\cite{Biggio:2003kp}.}  First, the gauge kinetic term
contains in its non-abelian part a quartic potential for the internal
components of the gauge field, and thus for the Higgs fields
\cite{Arkani-Hamed:2001nc}. This opens the possibility of
increasing the Higgs mass to acceptable values. Second, the gauge
symmetry allows for the appearance of an operator, localized at the
orbifold fixed points, that is proportional to the internal component
of the gauge field strength in the hypercharge direction and contains
a mass term for the Higgs fields in its non-abelian part \cite{Csaki:2002ur,vonGersdorff2}.
These tadpoles are generated with quadratically divergent coefficients and
can unfortunately destabilize the EWSB scale. In supersymmetric
models, they correspond to localized Fayet--Iliopoulos (FI) terms,
since the $4D$ vector auxiliary field $D$ is identified with its $6D$
counterpart shifted by the internal components of the gauge field
strength \cite{Arkani-Hamed:2001tb}.

In the light of the above remarks, it is of primary importance
to understand whether and to what extent the idea of gauge--Higgs
unification can be implemented with qualitative success in $D>5$
dimensions, before attempting to build a realistic model. One more
drawback of $D>5$ models with respect to $D=5$ models, besides
the possible occurrence of quadratic divergences, is some loss of
predictivity. Indeed, there are typically more geometric moduli,
parametrizing the shape and size of the internal space, and also
more Higgs fields, since there are more internal dimensions. The
simplest model of gauge--Higgs unification in $6D$ can be obtained
by considering an $SU(3)$ gauge theory on a $T^2/\Z_2$ orbifold
\cite{Antoniadis:2001cv}, and gives rise to three geometric moduli
and two Higgs doublets. In this model, there turns out to be a
parity symmetry that can forbid the appearance of the
divergent tadpole, or allow to control its size through some parameter
if it is softly broken. The Higgs potential in this model has the
same structure as that of the Minimal Supersymmetric Standard Model
(MSSM) and it seems again difficult to get reasonable masses for all
the Higgs fields after the EWSB.

The aim of this paper is to explore all $6D$ toroidal orbifold
constructions of the form $T^2/\Z_N$ (with $N=2,3,4,6$), giving
rise to $6D$ gauge--Higgs unification without SUSY. We mainly
focus on the minimal $SU(3)$ unified gauge symmetry, which is
broken to the SM $SU(2)\times U(1)$ EW symmetry by the orbifold
projection, one or more Higgs fields being responsible for
EWSB.\footnote{See \cite{higgsless} for the possibility of constructing
Higgless theories where EWSB is achieved by boundary conditions
and unitarity breaking occurs at scales higher than $m_Z$.}
Differently from the $N=2$ model, which necessarily leads to two
Higgs doublets, $N>2$ models offer more possibilities and can lead
for instance to a single Higgs doublet. Interestingly, in contrast
to the $5D$ case, one also gets a non-vanishing tree-level
quartic coupling, given by the usual gauge coupling. Under the
assumption that EWSB occurs, the Higgs mass in these models is therefore
predicted to be twice the $W$ mass at tree-level:
\begin{equation}
m_H = 2\, m_W \,.
\label{preintro}
\end{equation}
For these $N>2$ models, however, there seems to exist no symmetry able
to forbid the localized divergent tadpole, and the electroweak scale is
therefore expected to be unstable.

In the following, we present an explicit one-loop computation of the tadpole
coefficients and show that the corresponding operator is indeed radiatively
generated at the orbifold fixed points.
We study in detail the contributions of scalar, spinor and vector fields,
and show that even an accidental cancellation at each fixed point seems
impossible without introducing fundamental scalars.
On the other hand, the integral of
the tadpole over the compact space can happen to vanish. In this case,
one can expect that its presence should not affect the mass of the Higgs
field, as happens for a globally vanishing FI term in $5D$ SUSY models
\cite{Barbieri:2002ic}. A complete analysis of the effects of general
localized tadpoles
on the wave functions and on the spectrum of the Higgs modes is not totally
straightforward. Fortunately, the case of a globally vanishing tadpole
can be analysed along the lines of \cite{Lee:2003mc} for the SUSY case.
It turns out that a globally vanishing tadpole induces a non-trivial
gauge-field background that does not give rise to EWSB and in which there
indeed exists a zero-mode for the Higgs field. Its wave function has a
non-trivial profile along the compact space and displays localization
or delocalization at the fixed points, in complete analogy with SUSY
theories with localized FI terms \cite{Lee:2003mc}.

We believe that higher-dimensional orbifold constructions of this type,
with a single Higgs field, a tree-level quartic potential and a vanishing
integrated tadpole at the one-loop level, represent an extremely promising
class of models. EWSB can be induced by finite radiative corrections to
the Higgs mass term, associated to non-local operators, which we compute in the
following, and is stable at the one-loop level. A direct sensitivity of the
EW scale to the cut-off can only arise at two loops, and the little hierarchy
problem is solved.

The paper is organized as follows. In section 2 we introduce the $T^2/\Z_N$
orbifolds and analyse the possible $4D$ field configurations that can be
obtained.  In section 3 the classical and quantum forms of the Higgs potential
are studied. In section 4 we compute the contributions of bulk gauge fields,
as well as scalars and fermions in arbitrary representations, to the divergent
localized tadpole. Finally, section 5 contains a discussion of the effect
induced by the tadpole and general phenomenological implications.

\section{Orbifold models in six dimensions}

Let us consider a $6D$ gauge theory compactified on the orbifold $T^2/\Z_N$.
The $2D$ torus is parametrized by three real parameters, the two radii
$R_1$ and $R_2$ and an angle $\theta$, and is defined by identifying points
in a plane as
\begin{eqnarray}
y^1 \a\a \sim y^1 + n\, 2\pi R_1 + m\, 2\pi R_2 \cos \theta \,, \nn \\
y^2 \a\a \sim y^2 + m \, 2\pi R_2 \sin \theta\,,
\label{torodef}
\end{eqnarray}
for any integers $m$,$n$. It is useful to introduce complex coordinates
$z = \frac 1{\sqrt{2}}(y^1 + i y^2)$, so that the metric
components\footnote{Our convention for the 6D metric is mostly minus.}
with $A,B=z,\bar z$ are given by $g_{z \bar z} = g^{z \bar z} = -1$.
Defining the modular parameter $U = \frac {R_2}{R_1} e^{i \theta}$, and
renaming $R\equiv R_1$, the lattice (\ref{torodef}) can then be rewritten
in the complex plane as
\begin{eqnarray}
z \sim z + (m + n\, U)\frac {2\pi R}{\sqrt{2}} \,.
\end{eqnarray}

The generator $g$ of the orbifold group $\Z_N$ acts on the torus as a
$\frac {2\pi}N$ rotation. Consistent orbifold constructions are constrained
by the possible crystallographic symmetries of $2D$ lattices. They exist only
for $N=2$ with arbitrary $U$ and for $N=3,4,6$ with $U=e^{\frac{2 \pi i}N}$
or other equivalent discrete choices. This means that for $N=3,4,6$ there is
only one K\"ahler modulus parametrized by $R$, as in the $5D$ model on
$S^1/\Z_2$, whereas in the degenerate case $N=2$ there is in addition a
complex structure modulus $U$. The orbifold generator acts on the coordinates
as $z \rightarrow \tau z$, with $\tau=e^{\frac{2 \pi i}N}$, and is also
embedded into the gauge group through a matrix $P$ such that $P^N=1$.
For simplicity, we consider only group actions in the gauge sector that
correspond to inner automorphisms of the gauge group $G$ (see {\em e.g.}
\cite{Hebecker:2001jb} for a discussion of various orbifold gauge actions),
in which $P\in G$ (up to a constant overall phase for matter fields).

The Lagrangian of the orbifold theory is constrained to be the sum of a
bulk contribution, which must be invariant under the full gauge group,
and a set of contributions localized at the fixed points of the orbifold
action, which have to be invariant only under the gauge group surviving
at these points.
The set of points left fixed by an element $g^k$ of the orbifold group
depends on $k=0,1,\dots,N-1$, and it is therefore necessary to distinguish
sectors labelled by different $k$. Since $g^{N-k}$ is the inverse of
$g^{k}$, the fixed points in the sectors $k$ and $N-k$ are the same, and
their number is given by
\begin{equation}
N_k = \Bigg[ 2 \sin \bigg(\frac{\pi k}{N}\bigg) \Bigg]^2\,.
\label{Nk}
\end{equation}
Moreover, the sector $k=0$ is trivial and has of course no fixed points.
The physically distinct and relevant sectors are therefore labelled by
$k=1,\ldots, [N/2]$, where $[\dots]$ denotes the integer part. The general
form of the effective Lagrangian can therefore be parametrized as
\begin{equation}
{\cal L} = {\cal L}_6 + \sum_{k=1}^{[N/2]}
\sum_{i_k=1}^{N_k} \delta^{(2)}(z-z_{i_k}){\cal L}_{4,i_k} \,,
\label{LagLoc}
\end{equation}
where ${\cal L}_6$ represents the bulk $6D$ Lagrangian and ${\cal
L}_{4,i_k}$ the localized Lagrangians at the $N_k$ $g^k$ fixed points.
Since ${\cal L}$ has to be $g$-invariant, and $g$ acts non-trivially
on some fixed points, there are in general various non-trivial constraints
among the ${\cal L}_{4,i_k}$'s. Moreover, the orbifold structure respects
a discrete translational symmetry mapping $g$ fixed points onto $g$ fixed
points.\footnote{This is true only in the absence of localized matter that is
not uniformly distributed over the fixed points or of discrete Wilson lines.}
This implies that the Lagrangians ${\cal L}_{4,i_k}$ are constrained to be
all equal at fixed $k$ and hence there are only $[N/2]$ independent localized terms
appearing in (\ref{LagLoc}).

Contrary to the more familiar cases of the $\Z_2$ and $\Z_3$ orbifolds,
for the $\Z_4$ and $\Z_6$ orbifolds there are points that are fixed under
the action of some element $g^k$ of the group, but not fixed under some
subgroup of $\Z_N$, which permutes them. From eq.~(\ref{Nk}) one finds that the
$\Z_4$ orbifold has two $g$ (and $g^3$) fixed points and four $g^2$ fixed
points: the two $g$ fixed points, and two more points that are exchanged by
the action of $g$. The $\Z_6$ orbifold has one $g$ (and $g^5$) fixed point,
the origin $z=0$, three $g^2$ and four $g^3$ fixed points. Besides $z=0$,
the remaining two $g^2$ fixed points are exchanged by the action of
$g^3$, whereas the remaining three $g^3$ fixed points are exchanged by
the action of $g^2$. We summarize in Fig.~1 the orbifold fixed-point structure
of the $\Z_3$, $\Z_4$ and $\Z_6$ orbifolds, leaving aside the more familiar
$\Z_2$ case.

\begin{figure}[t]
\begin{center}
\begin{picture}(360,90)(0,-10)
\put(0,0){\begin{picture}(110,85)(0,0)
\SetColor{White}
\GBox(36.67,0)(73.34,63.51){0.8}
\begin{turn}{30}
\BBoxc(70,30)(50,23)
\BBoxc(60,-30)(50,23)
\end{turn}
\SetColor{Black}
\Vertex(36.67,0){3}
\Line(36.67,0)(110,0)
\Line(36.67,0)(0,63.51)
\Line(0,63.51)(73.34,63.51)
\Line(73.34,63.51)(110,0)
\Vertex(36.67,42.34){3}
\Vertex(73.33,21.17){3}
\ArrowLine(36.67,0)(36.67,42.34)
\ArrowLine(36.67,0)(73.33,21.17)
\ArrowLine(73.34,63.51)(36.67,42.34)
\ArrowLine(73.34,63.51)(73.33,21.17)
\Text(31,23)[c]{\footnotesize $A$}
\Text(51,56)[c]{\footnotesize $D$}
\Text(79,42)[c]{\footnotesize $C$}
\Text(60,8)[c]{\footnotesize $B$}
\end{picture}}
\put(125,0){\begin{picture}(110,85)(0,0)
\GBox(25,0)(55,30){0.8}
\Vertex(25,0){3}
\Vertex(55,30){3}
\Vertex(55,0){2}
\Vertex(25,30){2}
\Line(25,0)(85,0)
\Line(85,0)(85,60)
\Line(85,60)(25,60)
\Line(25,60)(25,0)
\ArrowLine(55,0)(55,30)
\ArrowLine(25,30)(25,0)
\ArrowLine(25,30)(55,30)
\ArrowLine(55,0)(25,0)
\Text(20,15)[c]{\footnotesize $A$}
\Text(40,35)[c]{\footnotesize $B$}
\Text(60,17)[c]{\footnotesize $C$}
\Text(40,-6)[c]{\footnotesize $D$}
\end{picture}}
\put(250,0){\begin{picture}(110,85)(0,0)
\SetColor{White}
\GBox(0,0)(36.67,63.51){0.8}
\begin{rotate}{30}
\BBoxc(25,-10)(50,20)
\begin{rotate}{30}
\BBoxc(45,19)(90,37)
\end{rotate}
\end{rotate}
\SetColor{Black}
\Vertex(0,0){3}
\Line(0,0)(73.34,0)
\Line(73.34,0)(110.01,63.51)
\Line(110.01,63.51)(36.67,63.51)
\Line(36.67,63.51)(0,0)
\Vertex(36.67,21.17){2}
\Vertex(73.33,42.34){2}
\Vertex(36.67,0){1}
\Vertex(18.33,31.76){1}
\Vertex(55,31.76){1}
\ArrowLine(0,0)(36.67,21.17)
\ArrowLine(0,0)(18.33,31.76)
\ArrowLine(36.67,63.51)(18.33,31.76)
\ArrowLine(36.67,63.51)(36.67,21.17)
\Text(2,19)[c]{\footnotesize $A$}
\Text(24,52)[c]{\footnotesize $D$}
\Text(43,44)[c]{\footnotesize $C$}
\Text(22,7)[c]{\footnotesize $B$}
\end{picture}}
\end{picture}
\caption{
\footnotesize Form left to right, the $T^2/\Z_3$, $T^2/\Z_4$
and $T^2/\Z_6$ orbifolds and their covering tori. We indicate
with points of decreasing size the $g$, $g^2$ and $g^3$ fixed points
respectively. The grey region represents the fundamental domain of the
orbifolds, and the segments delimiting it must be identified according to:
$A\sim D$, $B\sim C$.
}
\end{center}
\protect\label{fig:fp}
\end{figure}
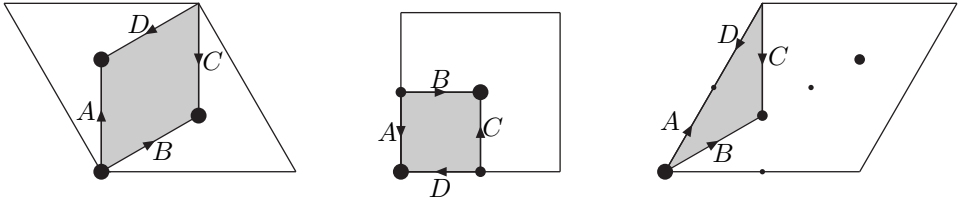

In the following we shall restrict our study to the prototype models of
gauge--Higgs unification with a gauge group $G=SU(3)$ that is broken to
$H=SU(2) \times U(1)$ by the $\Z_N$ orbifold projection. We denote by
$t^a$ the $SU(3)$ generators with the standard normalization ${\rm
Tr}\,t^a t^b = \frac 12 \delta^{ab}$ in the fundamental representation. The
unbroken generators in $SU(2)$ and $U(1)$ are $t^{1,2,3}$ and $t^8$. The broken
generators in $SU(3)/[SU(2) \times U(1)]$ are instead $t^{4,5,6,7}$, and can be
conveniently grouped into the usual raising and lowering combinations
$t^{\pm 1} = \frac 1{\sqrt{2}}(t^4 \pm i t^5)$ and $t^{\pm 2} = \frac
1{\sqrt{2}}(t^6 \pm i t^7)$.  In this basis, the group metric in the
sector $\pm i\,,\pm j$ is given by $h_{+ i\,, -j} = h^{+i\,,- j} =
\delta^{ij}$.  The most general way to realize the above breaking is
obtained by embedding the orbifold twist in the gauge group through
the matrix
\begin{equation}
P = \tau^{2\,n_p (\frac 13 + \frac 1{\sqrt{3}}t^8)} =
\left(\matrix{
\tau^{n_p} & 0 & 0 \cr
 0 & \tau^{n_p} & 0 \cr
 0 &  0 & 1 \cr}
\right)\;.\label{Rtwist}
\end{equation}
The number $n_p$ must be an integer and is defined only modulo $N$,
so that there are a priori $N-1$ inequivalent embeddings.

The geometric part of the $\Z_N$ action on a field is fixed by the
decomposition of its representation under the $6D$ $SO(1,5)$ Lorentz
group in terms of $SO(1,3)\times SO(2)$, where $SO(1,3)$ is the $4D$
Lorentz group and $SO(2)\simeq U(1)$ is the group of internal rotations.
The gauge part of the action on a field in a representation $\R$ of
$SU(3)$ is instead given by the twist matrix (\ref{Rtwist}) generalized
to the representation $\R$. This fixes the $\Z_N$ properties of any
field, up to an arbitrary overall phase $g$, such that the $N$-th power
of the $\Z_N$ action is trivial on all the components of the field.
The orbifold boundary condition of a generic bosonic or fermionic field
component $\Phi$, with $U(1)$ charge $s$ under internal rotations and in
the representation $\R$ of $SU(3)$, is then given by\footnote{Here and
in the following, for simplicity, we do not explicitly indicate the
dependence on $\bar z$ and on the $4D$ coordinates $x^\mu$.}
\begin{equation}
\Phi(\tau z) = g_{B,F}\, R_s \, P_\R \Phi(z) \,.
\label{bound_spin}
\end{equation}
In this equation, $P_\R$ denotes the twist matrix $P$ in the representation
$\R$ and $R_s=\tau^s$ is the Lorentz rotation associated to the geometric
action of the twist. The overall phases $g_{B,F}$ are such that $g_B^N=1$
for bosons and $g_F^N=-1$ for fermions, since $R_s^N = \pm 1$ in the two
cases. It is convenient to define $g_F = g \tau^{\frac 12} $, $g_B=g$, so that
$g$ is an $N$-th root of unity for both bosons and fermions. Correspondingly,
there are in general $N$ different boundary conditions, associated to the
$N$ possible choices of $g$. They are the $\Z_N$ analogues of the more
familiar even and odd parities appearing in $\Z_2$ models.

The expression of $P_\R$ can be
conveniently written as
\begin{equation}
P_\R= \tau^{2\, n_p (\frac{n_\R}3 + \frac 1{\sqrt{3}} t^{8}_\R)} \,,
\label{Pierre}
\end{equation}
where $t^{8}_\R$ is the Cartan generator $t^8$ in the representation $\R$
and $n_\R$ is an integer number such that $P_\R^N=1$. It can be written as
$n_\R=n_1-n_2$, where $n_1$ and $n_2$ are the two Dynkin labels of the
representation $\R$. Since the canonically normalized abelian generator
surviving the projection is $Q_\R = \frac 1{\sqrt{3}}t^8_\R$, the matrix
(\ref{Pierre}) gives a phase $\tau^{2\, n_p (\frac{n_\R}3 + q)}$ on a
component with $U(1)$ charge $q$ under the decomposition of the representation
$\R$ under $SU(3) \rightarrow SU(2) \times U(1)$. The relevant information
is listed in Table~\ref{decomp} for the first few representations. In the
following two subsections, we consider in some more detail the decomposition
of gauge and matter fields, as given by (\ref{bound_spin}).

\begin{table}[htb]
\begin{center}
\begin{tabular}{|c|l|c|}
\hline
$\R$ \b & \hfil{Decomposition of $\R$} & $n_\R$  \\
\hline
${\bf 3}$ \b
& ${\bf 2}_{\frac 16} \oplus {\bf 1}_{-\frac 13}\,$
& $1$ \\
${\bf 6}$ \b
& ${\bf3}_{\frac 13} \oplus {\bf 2}_{-\frac 16} \oplus {\bf 1}_{-\frac 23}\,$
& $2$ \\
${\bf 8}$ \b
& ${\bf 3}_0 \oplus {\bf 2}_{\frac 12} \oplus {\bf 2}_{-\frac 12}
\oplus {\bf 1}_0\,$
& $0$ \\
${\bf 10}$ \b
& ${\bf 4}_{\frac 12} \oplus {\bf 3}_{0} \oplus {\bf 2}_{-\frac 12}
\oplus {\bf 1}_{-1}\,$
& $3$ \\
\hline
\end{tabular}
\end{center}
\caption{\footnotesize Decomposition of the most relevant $SU(3)$
representations.}
\label{decomp}
\end{table}

\subsection{Gauge fields}

The gauge fields $A_M$ transform as vectors under $SO(1,5)$ rotations
and in the adjoint representation under gauge transformations. In
complex coordinates, the decomposition of $A_M$ under $SO(1,3)\times
U(1)$ is very simple: we get a $4D$ vector field $A_\mu$ with
charge $s=0$ and two $4D$ scalars $A_z$ and $A_{\bar z}$ with charges
$s=-1$ and $s=1$ respectively. The boundary conditions can be obtained from
eq.~(\ref{bound_spin}) with $g=1$. The gauge part of the orbifold
twist is diagonal if one switches from the standard basis with components
$A_{Ma}$ to the creation--annihilation basis with components $A_{M1,2,3,8}$,
$A_{M\pm 1} = \frac 1{\sqrt{2}}(A_{M4} \mp i A_{M5})$ and
$A_{M\pm 2} = \frac 1{\sqrt{2}}(A_{M6} \mp i A_{M7})$.
The final result is that the various components of the gauge field
$A_M = \sum_a A_{Ma}\,t^a$ satisfy twisted boundary conditions with
the following phases:
\begin{eqnarray}
\a\a A_{\mu 1,2,3,8} : 1 \,,\;
A_{z 1,2,3,8} : \tau^{-1} \,,\;
A_{\bar z\,1,2,3,8} : \tau^{+1} \,, \\
\a\a A_{\mu \pm i} : \tau^{\pm n_p} \,,\;
A_{z \pm i} : \tau^{-1 \pm n_p} \,,\;
A_{\bar z \pm i} : \tau^{+1 \pm n_p} \,.
\end{eqnarray}

The light modes of untwisted fields consist of the gauge bosons
$A_{\mu 1,2,3,8}$ forming the adjoint of the surviving gauge group,
the scalar fields $A_{z + i}$ with their complex conjugates
$A_{\bar z - i}$ forming a charged Higgs doublet under this group
if $n_p = 1$ mod $N$, and the scalar fields $A_{z -i}$ with their
complex conjugate $A_{\bar z + i}$ forming a conjugate charged Higgs
doublet if $n_p = -1$ mod $N$. Referring to the decomposition reported
in Table~\ref{decomp}, the projection keeps the ${\bf 3}_0$ and
${\bf 1}_0$ components for $4D$ indices and some numbers $n$ and $n_c$
of the ${\bf 2}_{\frac 12}$ and the ${\bf \bar 2}_{- \frac 12}$
components for internal indices, depending on $N$ and $n_p=1,\dots,N-1$.
The possibilities for the numbers $(n,n_c)$ for the consistent constructions
labelled by the integers $(N,n_p)$ are the following:
\begin{eqnarray}
(n,n_c) = (1,1) \a:\a \;{\rm for}\; (N,n_p)=(2,1)\,; \\
(n,n_c) = (1,0) \a:\a \;{\rm for}\; (N,n_p)=(3,1),(4,1),(6,1)\,; \\
(n,n_c) = (0,1) \a:\a \;{\rm for}\; (N,n_p)=(3,2),(4,3),(6,5)\,; \\
(n,n_c) = (0,0) \a:\a \;{\rm for}\; (N,n_p)=(4,2),(6,2),(6,3),(6,4)\,.
\end{eqnarray}
It is therefore possible to construct models with two conjugate
Higgs doublets ($\Z_2$), a single Higgs doublet ($\Z_3$, $\Z_4$, $\Z_6$)
or no Higgs doublets at all ($\Z_4$, $\Z_6$).

\subsection{Matter fields}

A $6D$ Weyl fermion $\Psi_\pm$ of definite $6D$ chirality decomposes under
$SO(1,3)\times U(1)$ into two $4D$ chiral fermions with charges
$s=\pm \frac{1}{2}$: $\Psi_\pm = (\psi_{L,R})_{s=\frac{1}{2}} \oplus
(\chi_{R,L})_{s=-\frac{1}{2}}$, where $L,R$ denote the $4D$ chiralities.
We thus see from (\ref{bound_spin}) that any $6D$ Weyl spinor gives rise
to two $4D$ fermions of opposite $4D$ chiralities, twisted by $g$ and $g\tau$,
times the gauge part of the twist. More generally, a $6D$ spinor field
$\Psi_{\R,\chi_6}$ of $6D$ chirality $\chi_6 = \pm 1$ transforming in a
representation $\R$ of the gauge group, gives rise to different $4D$ spinor
components $\psi_{q,\chi_4}$ with $U(1)$ charge $q$ and $4D$ chirality
$\chi_4 = \pm 1$, twisted by a phase:
\begin{equation}
\psi_{q,\chi_4} : g\,\tau^{\frac {1-\chi_4\chi_6}2}
\tau^{2\, n_p (\frac{n_\R}3 + q)} \,.
\end{equation}
Depending on $N$ and $n_p$, the various possible choices for $g$ allow
the zero modes of different subsets of components to be preserved.
We will not list here the many possibilities, since they can be easily
derived from the data reported in Table~\ref{decomp}.

For scalar fields the analysis is simpler, since they are singlets
under Lorentz transformations and thus $s=0$ in (\ref{bound_spin}).
The twist of a scalar field $\phi_\R$ in a representation $\R$ of
the gauge group is only given by its gauge decomposition.
For a generic component $\phi_{q}$ with $U(1)$ charge $q$, one has
\begin{equation}
\phi_{\R,q} : g\, \tau^{2\, n_p (\frac{n_\R}3 + q)} \,.
\end{equation}
Notice that there is a one-to-one correspondence between the case of
scalars and that of spinors, since the additional phase
$\tau^{\frac {1-\chi_4\chi_6}2}$ arising for the latter is always
an $N$-th root of unity and can therefore be compensated by a different
choice of $g$. It is easy to verify that the zero mode of any component
can always be preserved with a suitable and unique choice of the phase $g$,
both for scalars and for fermions. This is an important property for model
building.

\subsection{Wave functions and spectrum}

To construct wave functions, it is convenient to introduce two
alternative real coordinates $w_1$ and $w_2$, which are aligned
with the natural cycles specified by the complex structure
$U = U_1 + i U_2$ and defined by the relation
$z= \frac 1{\sqrt{2}}(w_1 + U w_2)$. In this way, $w_1$ and
$w_2$ are independently periodic with period $2 \pi R$. For $N=3,4,6$,
where $U=\tau$, the $\Z_N$ twist changes the point $(w_1,w_2)$ into the point
$(-w_2,w_1 + 2 \tau_1 w_2)$, where $\tau_{1,2}$ denotes the
real and imaginary parts of $\tau$.
For $\Z_2$, one has simply $(w_1,w_2)\rightarrow \tau(w_1,w_2)$.
It will be convenient in the following
to introduce a matrix notation, in which the vector $\vec w$ is
transformed into the vector $Z_N^t \vec w$. The matrix $Z_N$ is
given by
\begin{equation}
Z_N = \left(\matrix{0 & 1 \cr -1 & 2 \tau_1}\right)
\label{zeta}
\end{equation}
for $N=3,4,6$, while $Z_2=-I$.
The basis of periodic functions on $T^2$ is then given by the usual
exponential functions $f_{\vec n}(\vec w) \sim e^{\frac iR \vec n
\cdot \vec w}$.  In terms of the complex variable $z$, the normalized
result is
\begin{equation}
f_{\vec n}(z) = \frac{1}{\sqrt{V}} e^{\frac 1{\sqrt{2}}
(\lambda_{\vec n} z - \bar \lambda_{\vec n} \bar z)} \,,
\label{f}
\end{equation}
where $V$ is the volume of the covering torus and
\begin{equation}
\lambda_{\vec{n}} =\frac {n_2 - n_1 \bar U}{U_2\, R} \,,\;\;
\bar\lambda_{\vec{n}}=\frac {n_2 - n_1 U}{U_2\, R} \;.
\label{lambda}
\end{equation}
The $\Z_N$ twist acts on $f_N$ and $\lambda_{\vec n}$ as
\begin{equation}
f_{\vec n}(\tau^k z) = f_{Z_N^k \vec n}(z) \,,\;\;
\lambda_{Z_N^k \vec n} = \tau^k \lambda_{\vec n} \;.
\label{ftransf}
\end{equation}
It is easy to construct $\Z_N$ covariant wave functions on $T^2$
by applying to the functions (\ref{f}) the orbifold projection weighted
by an arbitrary $\Z_N$ phase $g$. Defining for convenience the quantity
$\eta_{\vec n} = \left(\sqrt{N}\right)^{-\delta_{\vec n,\vec 0}}$,
these are given by
\begin{equation}
h_{\vec n}^g(z) = \frac {\eta_{\vec n}}{\sqrt{N}}
\sum_{k=0}^{N-1} g^{-k} f_{Z_N^k \vec n}(z) \;,
\end{equation}
and, thanks to (\ref{ftransf}), satisfy the generic twisted boundary condition
$h^g_{\vec n}(\tau z) = g h^g_{\vec n}(z)$. It is easy to verify that these
functions are also orthonormal with respect to the Kaluza--Klein (KK)
momenta as well as the twist $g$.
However, the functions $h_{\vec n}^g(z)$ are not all independent:
those with mode vectors connected by the orbifold action are proportional
to each other through a phase:
\begin{equation}
h_{Z_N^k \vec n}^g(z) = g^{k} h_{\vec n}^g(z) \;.
\label{rel}
\end{equation}
Correspondingly, the mode vectors $\vec n$ are not all independent but
restricted to belong to some fundamental domain, which can be
determined as follows.  The matrix (\ref{zeta}) represents the $\Z_N$
action on the mode vector $\vec n$ for the torus wave functions.
For $N\neq 2$, it amounts to a rotation with phase $\tau$ on
the complex plane $u = - n_1 + \tau n_2$.
This means that we can divide the space $\ZZ^2$ of all possible mode
vectors $\vec n$ into the origin, which is left fixed by $Z_N$, plus
$N$ sectors $D_k$, with $k=0,\dots,N-1$, mapped into each other by $Z_N$.
For $N>2$, these domains can all be defined as $D_k = \{\vec n \in \ZZ^2|
(Z_N^{k} \vec n)_1 < 0, (Z_N^{k} \vec n)_2 \ge 0\}$, whereas for $N=2$,
they are given by $D_0=\{\vec n \in \ZZ^2 | n_1>0 \oplus (n_1=0,n_2>0)\}$,
$D_1=\{\vec n \in \ZZ^2| n_1<0 \oplus (n_1=0,n_2<0)\}$.
The independent wave functions in (\ref{rel}) are then associated to
$\vec n \in D_0$ and the origin, the ones associated to $\vec n \in D_{k}$
with $k \neq 0$ being the $\Z_N$-transformed of these.

It is now straightforward to characterize the spectrum of a generic $T^2/\Z_N$
orbifold model. A field $\phi^g(z)$ with generic twisted boundary conditions
\begin{equation}
\phi^g(\tau z) = g \phi^g(z)
\end{equation}
can be expanded in KK modes as
\begin{equation}
\phi^g(z) = \delta^{g,1} \phi_{\vec 0}^1 h_{\vec 0}^1(z)
+ \sum_{\vec n \in D_0} \phi_{\vec n}^g h_{\vec n}^g(z) \;.
\label{PhiKK}
\end{equation}
The mass $m_{\vec n}$ of the $\vec n$-mode is given by
\begin{eqnarray}
m_{\vec n} = |\lambda_{\vec n}| =
\frac {\sqrt{n_1^2 + n_2^2 - 2 U_1 n_1 n_2}}{U_2 R}\,.
\end{eqnarray}
It is important to notice that the spectrum of modes does not depend
on $g$, apart from the zero mode, which exists only if $g=1$.

\section{Higgs potential}

The biggest problem in achieving gauge--Higgs unification in the
minimal $5D$ case is the absence of a tree-level Higgs potential,
resulting in too small a Higgs mass. This is the main reason for
considering gauge--Higgs unification in $6D$, where such tree-level
quartic term, arising from the gauge kinetic term, is naturally
present. As suggested by several authors
\cite{Arkani-Hamed:2001nc,Antoniadis:2001cv}, its presence can help
getting realistic EWSB and Higgs masses. Most of the $6D$ models
discussed so far, however, were based on $\Z_2$ orbifold constructions
that necessarily lead to two charged Higgs doublets. In this case,
the tree-level quartic term has a flat direction, just as in the MSSM,
and therefore fluctuations along this direction only have radiatively
induced masses, which in general tend to be too small.

We now focus our attention on $T^2/\Z_N$ orbifold constructions with
$N>2$ leading to one Higgs doublet. As we shall show below, these models
have a non-vanishing quartic tree-level potential, in contrast to the
$S^1/\Z_2$ orbifold. This term is responsible for an important distinction
between the interpretation of EWSB in $T^2/\Z_N$ and $S^1/\Z_2$ orbifolds.
In the $5D$ model, the vacuum expectation value (VEV) of the Higgs field
is a flat direction of the classical potential and corresponds to a Wilson
loop, which is also equivalent to a twist in the boundary conditions around
$S^1$ \cite{hos}. In $6D$ models, on the contrary, the VEV of the Higgs field
is {\it not} a flat direction of the classical potential, and such
interpretation is missing. Indeed, there exist no continuous families of
solutions to the usual orbifold consistency conditions for Wilson loops
\cite{Dixon} in the case of $SU(3)$ gauge theories on $T^2/\Z_N$ with $N>2$.
Only discrete Wilson loops are allowed. Nevertheless, the $5D$ and $6D$ models
share the interesting property that the Higgs dynamics is much more
constrained than what is just implied by the surviving gauge symmetry. This is
a consequence of the non-linearly realized remnant of the higher-dimensional
gauge symmetry associated to parameters depending on the internal coordinates,
under which the Higgs field transforms inhomogeneously \cite{vonGersdorff1}.

Let us now compute the classical Higgs potential that arises for the single
Higgs models on $T^2/\Z_N$ with $N=3,4,6$. We choose $n_p=1$, but the case
$n_p=N-1$ is perfectly similar up to an overall conjugation and therefore
physically equivalent. The classical Lagrangian of the $6D$ theory is given
simply by $L=-\frac 12 {\rm tr} F_{MN}^2$, where
$F_{MN} = \partial_M A_N - \partial_N A_M - i g_6 \left[A_M,A_N\right]$.
The Lagrangian for the zero modes $A_\mu^0$, $A_{z}^0$ and $A_{\bar z}^0$
is easily obtained by integrating over the internal torus. The result
is given by
\begin{equation}
L = -\frac 12\,{\rm tr}\,F_{\mu\nu}^{0 2}
+ 2\, {\rm tr}\,|D_\mu A_z^0|^2
- g_{4}^2\, {\rm tr}\,[A_{z}^0, A_{\bar z}^0]^2\,,
\end{equation}
where $g_4 = g_6/\sqrt{V}$ is the gauge coupling of the $4D$ effective
theory below the compactification scale,
$F_{\mu \nu}^0 = \partial_\mu A_\nu^0 - \partial_\nu A_\mu^0
- i g_4 [A_\mu^0,A_\nu^0]$ is the field strength of the massless $4D$
gauge bosons, and $D_\mu A_{z,\bar z}^0 = \partial_\mu A_{z,\bar z}^0
- i g_4 [A_\mu^0,A_{z,\bar z}^0]$ is the covariant derivative on the Higgs
field. The three weak gauge bosons and the hypercharge gauge boson are
identified as $W_{\mu a} = A_{\mu a}^0$ for $a=1,2,3$ and
$B_{\mu} = A_{\mu 8}^0$. The zero modes of $A_\mu = \sum_a A_{\mu a} t^{a}$,
where $a=1,2,3,8$, are then given by
\begin{equation}
A_{\mu}^0 = \frac 12
\left(\matrix{
W_\mu^3 + \frac 1{\sqrt{3}}\, B_\mu & \sqrt{2} W_\mu^+ & 0 \cr
\sqrt{2} W_\mu^- & -\,W_\mu^3 + \frac 1{\sqrt{3}}\,B_\mu & 0 \cr
0 & 0 & - \frac 2{\sqrt{3}}\, B_\mu \cr}
\right)\;.
\end{equation}
Similarly, the two complex components of the Higgs doublet are
$h_u = A_{z +1}^0$ and $h_d = A_{z +2}^0$, and their complex conjugates
are given by $h_u^* = A_{\bar z -1}^0$ and $h_d^* = A_{\bar z -2}^0$.
The zero modes of $A_z = \sum_i A_{z+i} t^{+i}$ and
$A_{\bar z} =\sum_i A_{\bar z-i} t^{-i}$ are thus given by
\begin{equation}
A_{z}^0 = \frac{1}{\sqrt{2}}
\left(\matrix{
0 & 0 & h_u \cr
0 & 0 &  h_d \cr
0 & 0 & 0 \cr}
\right)\;,\;\;
A_{\bar z}^0 =\frac{1}{\sqrt{2}}
\left(\matrix{
0 & 0 & 0 \cr
0 & 0 & 0 \cr
h_u^* & h_d^* & 0 \cr}
\right)\;.
\label{AzAzbar}
\end{equation}

Substituting these expressions in the Lagrangian, and switching from
the $SU(3)$ to an $SU(2)$ notation, we finally find:
\begin{eqnarray}
L = - \frac 12\, {\rm tr}\,F_{\mu\nu}^{W2} - \frac 14 F_{\mu\nu}^{B2}
+  \Big|\Big(\partial_\mu - i g_4 W_{\mu a} \frac {\tau_a}2
- i g_4 \tan \theta_W \frac 12 B_\mu\Big)h\Big|^2 - V_{\rm class}(h)\,,
\end{eqnarray}
where $\tan \theta_W = \sqrt{3}$ and
\begin{eqnarray}
V_{\rm class}(h) \a=\a \frac {g_{4}^2}2 \big|h\big|^4 \;.
\label{classical-pot}
\end{eqnarray}
The weak mixing angle arising in this construction is too large, but there
are various ways of solving this problem, most notably by adding extra
$U(1)$ gauge fields.

Quantum fluctuations induce a correction to the classical potential
(\ref{classical-pot}) and can trigger radiative symmetry breaking.
The quantum effective potential can only depend on gauge-invariant quantities.
These can be local or non-local in the compact dimensions.
Non-local operators involve Wilson lines wrapping around the internal space
and are generated with finite coefficients whose size is controlled by the
compactification scale $1/R$. The local and potentially divergent operators
contributing to the Higgs potential arise from the non-derivative part
of $F_{z \bar z}$, like the classical quartic term. Gauge invariance allows
two possible classes of local operators of this kind: even powers of
$F_{MN}$ in the bulk or arbitrary powers of $F_{z\bar z}$ localized at the
orbifold fixed points. In general, such terms will be generated at the
quantum level with divergent coefficients. At one-loop order, the bulk
operators that can lead to divergences in the Higgs potential are the gauge
kinetic term $F_{MN}^2$ and a quartic coupling $F_{MN}^4$, leading
to quadratic and logarithmic divergences to $V(h)$ respectively.
Localized operators are of the form $g_4^{2p} F_{z \bar z}^p$, where $p$ is
any positive integer. Quadratic and logarithmic divergences can arise
from the tadpole operator $p=1$ and the kinetic operator $p=2$ respectively.

Since the quadratic bulk divergence gives rise only to a wave-function
renormalization, we see that the only quadratic divergence to the Higgs
potential comes from the localized tadpole operator $F_{z\bar z}$.
In general, the latter induces a modification to the background and,
in its non-abelian part, possible mixings between the Higgs and its KK modes,
aside from a quadratically divergent mass term for the Higgs field $h$.
In the rough approximation of neglecting the backreaction induced by the
modified background and the KK mixings, effects that we will consider in
section 5, and also neglecting all the logarithmic divergences, we see
that the leading terms in the one-loop effective potential for the Higgs
are
\begin{eqnarray}
V_{\rm quant}(h) = - \mu^2 |h|^2 + \lambda |h|^4 \,,
\label{Vheff}
\end{eqnarray}
where $\mu^2$ is a radiatively generated and possibly divergent mass
term and $\lambda = g^2/2$ is the tree-level quartic term.
Assuming $\mu^2>0$ so that EWSB can occur, we have $\langle |h|
\rangle = v/\sqrt{2}$ with $v = \mu/\sqrt{\lambda}$.
At the minimum,
\begin{eqnarray}
m_H = \a\a \sqrt{2}\, \mu =  \sqrt{2}\,v\,\sqrt{\lambda} \nn \\
m_W = \a\a \frac12 \, g\, v \,.
\label{mh-mw}
\end{eqnarray}
The ratio between $m_H$ and $m_W$ is therefore predicted in a
completely model-independent way to be
\begin{equation}
\frac {m_H}{m_W} = \frac {2\sqrt{2\lambda}}{g} = 2 \,.
\label{pred}
\end{equation}
Extra $U(1)$ fields, possibly needed to fix the weak-mixing angle to
the correct value, do not modify eq.~(\ref{pred}).
The main radiative correction to eq.~(\ref{pred}) arises from the
Higgs wave-function distortion induced by the tadpole operator
$F_{z\bar z}$, as explained in section 5. This effect can be estimated
by Na\"{\i}ve Dimensional Analysis (NDA) to give $O(1)$ corrections
to eq.~(\ref{pred}). In spite of this, the value of the Higgs mass is
significantly increased with respect to the previously considered
$5D$ models or $\Z_2$ orbifold constructions.

\section{Divergent localized tadpole}

We have seen that gauge invariance allows a localized interaction that is
linear in the field strength, in addition to the universally allowed
higher-order interactions involving even powers of the field strength.
The localized interaction is particularly relevant, since it involves a
mass term for the Higgs fields \cite{vonGersdorff2}. It has the
form\footnote{Abelian gauge fields that are present already before
the orbifold projection and are unbroken can also develop a localized
divergence term as in (\ref{O}), but in this case the associated divergent
mass term for even scalars, the last factor in (\ref{tadpolo}), is absent.}
\begin{equation}
{\mc L}_{\rm tad} =- i \sum_{k=1}^{[N/2]} \frac {\mc C_k}{N_k}
\sum_{i_k=1}^{N_k} \delta^{(2)}(z-z_{i_k}) F^8_{z \bar z}(z)\,,
\label{O}
\end{equation}
where $\mc C_k$ are real coefficients of mass dimension 1 and
$F_{z \bar z}^8$ is the field strength of the $U(1)$ component
left unbroken by the orbifold breaking, which in terms
of $6D$ fields reads
\begin{equation}
F_{z \bar z}^8 = \partial_z A_{\bar z}^8 - \partial_{\bar z} A_{z}^8
+ g_6 f^{8bc}A_{z\,b} A_{\bar z\,c}\,.
\label{tadpolo}
\end{equation}
In $\Z_2$ orbifold models, the parity symmetry $z\leftrightarrow \bar z$ can
be implemented and it forbids the appearance of the operator (\ref{tadpolo}),
which is odd under this discrete symmetry \cite{Csaki:2002ur,vonGersdorff2}.
This parity can be generalized to $\Z_N$ orbifolds, with $N>2$, only if the
twist matrix $P$ is such that $P^2=I$. The allowed form of the tadpole
operator is then ${\rm Im}\,{\rm Tr}\, P F_{z \bar z}$, which
automatically vanishes whenever $P^2=I$. More precisely, we will see
that the term associated to $k$ in (\ref{O}) can be written as
${\rm Im}\,{\rm Tr}\, P^k F_{z \bar z}$, implying that the tadpole vanishes
in the sectors $k$ such that $P^{2k}=I$, when the above $\Z_2$ symmetry can
be implemented. Notice that projections that leave only one Higgs doublet
do not satisfy $P^2=I$ and hence are generally affected by tadpoles.
We verify this statement by performing a detailed calculation of the
coefficients ${\mc C_k}$ for all $\Z_N$ models at one-loop order.
In particular we compute the contribution to the tadpole arising
from gauge (and ghost) fields, and from an arbitrary bulk scalar or fermion
in a representation ${\mc R}$ of $SU(3)$. Possible localized boundary fields
cannot minimally couple to the fields appearing in (\ref{tadpolo}), because
of the residual non-linearly realized gauge symmetries that are unbroken
at the orbifold fixed points \cite{vonGersdorff1}. We therefore consider
in the following $6D$ bulk fields only. This computation is also useful
to understand whether and under what circumstances an accidental one-loop
cancellation is possible.

The computation of Feynman diagrams on an orbifold can be nicely mapped
to that on the corresponding covering torus by using a mode decomposition
such that the effect of the orbifold projections amounts only to a
non-conservation of the KK momentum in \emph{non-diagonal}
propagators,\footnote{An alternative procedure
to compute Feynman diagrams on orbifolds is obtained by directly considering
physical modes only, as derived in section 2. In this case, the propagators
for all fields are diagonal in the KK momenta and the momentum
non-conservation arises from the interaction vertices. As a further
consistency check of our results, we have also computed both the 1- and
2-point functions in this way and found perfect agreement between the
two methods.} along the lines of \cite{Georgi:2000ks}. Let us illustrate
the formalism
for a generic field $\Phi (z)$ on a $T^2/\Z_N$ orbifold. The $\Z_N$ group
acts on $\Phi$ as described in (\ref{bound_spin}), through the operator
$\mc{P}= g R_s P_\R$, on both Lorentz and gauge indices: $\Phi(\tau z)={\mc
P}\Phi(z)$.  We define the mode expansion of $\Phi$ as
\begin{eqnarray}
\Phi(z)\equiv \sum_{\vec n}\Phi_{\vec n}f_{\vec{n}}(z)\,,
\label{FIexp}
\end{eqnarray}
where $f_{\vec{n}}$ is the basis of functions on $T^2$ defined in (\ref{f}).
The KK modes in (\ref{FIexp}), contrary to those appearing in (\ref{PhiKK}),
provide a redundant parametrization of $\Phi$, since they are not all
independent, owing to the condition (\ref{bound_spin}). Their propagators
will then be non-diagonal in the KK momentum space.

It is convenient to express $\Phi$ in terms of an unconstrained field
$\tl\Phi$ on $T^2$ with the same quantum numbers as $\Phi\,$, so that
(\ref{bound_spin}) is automatically satisfied:
\begin{equation}
\Phi(z)=\frac 1N
\sum_{k=0}^{N-1} {\mc P}^k \tl{\Phi}(\tau^{-k}z)\,.
\label{FIpar}
\end{equation}
The propagator of $\Phi$ can then be written in terms of the propagator
of $ \tl \Phi$ as\footnote{We write the
propagator in the form of a correlator among $\Phi$ and its
hermitian conjugate $\Phi^\dag$, but our formalism clearly applies
to real fields as well.}
\begin{eqnarray}
\langle \Phi(z_1)\Phi^\dag(z_2) \rangle \a=\a
\frac 1{N^2}\sum_{k,l=0}^{N-1} {\mc P}^k
\langle \tl\Phi(\tau^{-k}z_1) \tl\Phi^\dag(\tau^{-l}z_2) \rangle
({\mc P}^\dag)^l \nn \\
\a=\a \frac 1{N}\sum_{k=0}^{N-1} {\mc P}^k
\langle \tl\Phi(\tau^{-k}z_1) \tl\Phi^\dag(z_2) \rangle \, ,
\label{FIprop0}
\end{eqnarray}
where the last simplification in the above expression is a consequence
of the fact that the transformation
$\tl\Phi(z) \, \rightarrow \, {\mc P}^{-k}\tl\Phi(\tau^k z)$ is by assumption
a symmetry of the action. The propagator of $\tl\Phi$ is the standard
propagator on the torus and can be written as
\begin{eqnarray}
\langle \tl\Phi(z_1) \tl\Phi^\dag(z_2) \rangle\equiv
\sum_{\vec n} \tl G_{\vec n}\,f_{\vec n}(z_1-z_2) \nn \,,
\end{eqnarray}
where $\tl G_{\vec n}$ denotes the standard form of the propagator in
momentum space, and the torus periodicity conditions result only in
the quantization of the KK momenta in the internal directions.
Recalling that the orbifold action on the KK momenta is given
by (\ref{ftransf}), we then find
\begin{eqnarray}
\langle \Phi(z_1)\Phi^\dag(z_2) \rangle
=\frac 1N \sum_{k=0}^{N-1}\!\sum_{\vec n} {\mc P}^k\,\tl G_{\vec n}\,
f_{Z_N^{-k}\vec n}(z_1) f_{\vec n}^\dag (z_2)\, ,
\label{FIprop1}
\end{eqnarray}
whose Fourier transform is
\begin{eqnarray}
\langle \Phi_{\vec m}{\Phi_{\vec n}}^\dag \rangle=\frac 1N
\sum_{k=0}^{N-1} {\mc P}^k\tl G_{\vec n}\,
\delta_{\vec m,\,Z_N^{-k}\vec n}\; .
\label{FIprop}
\end{eqnarray}

\begin{figure}[t]
\begin{center}
\begin{picture}(400,65)(-30,5)
\put(30,0){\begin{picture}(160,60)(0,0)
\ArrowLine(0,30)(80,30)
\ArrowLine(80,30)(160,30)
\BCirc(80,30){7}
\LongArrow(33,37)(47,37)
\LongArrow(113,37)(127,37)
\Text(40,47)[c]{ $\vec n$}
\Text(120,47)[c]{ $Z^{-k}_N\vec n$}
\Text(40,16)[c]{$B$}
\Text(120,16)[c]{$A$}
\Text(80,30)[c]{$k$}
\end{picture}}
\Text(205,30)[c]{$=$}
\Text(255,30)[c]{$
\displaystyle{\frac{1}{N}} \left[{\mc P}^k\cdot \tl G_{\vec n} \right]_
{A\,B}$}
\end{picture}
\caption{\footnotesize Feynman rule for the propagators on an orbifold.
In the figure, $A,\,B$ are a generic set of indices labelling the state
and $k=0,\ldots,\,N-1$ is the possible twist of the propagator.}
\end{center}
\protect\label{PropFey}
\end{figure}

Equation (\ref{FIprop}) shows that the propagator on an orbifold can be
written as the sum of $N$ propagators, of which all but the first
violate momentum conservation. Any internal line of a Feynman
diagram is then the sum of the ``$k$'' propagators shown in Fig.~2,
in which an incoming momentum $\vec n$ is changed into an outgoing one
$Z_N^{-k}\vec n$. When using the Feynman rule shown in Fig.~2,
an orientation of the propagator is needed so as to distinguish
incoming and outgoing lines. If the field is complex this orientation
is naturally provided; if it is real, one orientation has to be chosen
to apply the rule of Fig.~2, but clearly the result does not depend on
this choice.

\begin{figure}[t]
\begin{center}
\begin{picture}(400,60)(-25,15)
\put(222,-5){\begin{picture}(150,50)(0,0)
\Vertex(75,25){3}
\ArrowLine(10,50)(75,25)
\ArrowLine(0,25)(75,25)
\ArrowLine(150,25)(75,25)
\ArrowLine(34,60)(75,25)
\ArrowLine(150,40)(75,25)
\DashCArc(75,25)(30,20,130){2}
\CBoxc(13,52)(17,7){White}{White}
\Text(37.5,18)[c]{\footnotesize ${\vec n}_1,\, A_1$}
\Text(8,54)[c]{\footnotesize ${\vec n}_2,\, A_2$}
\Text(112.5,18)[c]{\footnotesize ${\vec n}_K,\, A_K$}
\end{picture}}
\Text(-35,30)[l]{$ V={\mc V}_{\{A_1,\,{\vec n}_1\},\ldots,\,
\{A_K,\,{\vec n}_K\}}
\Phi_{\{A_1,{\vec n}_1\}}\ldots
\Phi_{\{A_K,{\vec n}_K\}}$}
\Text(200,30)[c]{\large $\sim$}
\put(30,0){\begin{picture}(160,60)(0,0)
\end{picture}}
\end{picture}
\caption{\footnotesize The diagrammatic representation of a generic effective
vertex $V$.}
\end{center}
\end{figure}

In this approach, all the interaction vertices conserve the KK momenta
and any diagram can thus be computed by simply applying the usual
Feynman rules, inserting the orbifold propagator as shown in Fig.~2.
The computation is, however, simplified by noting that any interaction
vertex has to be $\Z_N$-invariant. Its action on a set of $K$ fields,
with modes $\Phi_{\vec n_f}$ ($f=1,\ldots, K$), is
$\Phi_{\vec n_f}\rightarrow{\mc P}_{f}^k\Phi_{Z^k_{N}\vec n_f}$.
This leads to the following relation, valid for any interaction vertex
${\mc V}$ (see Fig.~3):
\begin{equation}
{\mc V}_{\{B_1,\,Z^{-k}_N{\vec n}_1\},\ldots,\,
\{B_K,\,Z^{-k}_N{\vec n}_K\}}\,
\left[{\mc P}_{1}^k\right]_{B_1 A_1}\ldots
\left[{\mc P}_{K}^k\right]_{B_K A_K}=
{\mc V}_{\{A_1,\,{\vec n}_1\},\ldots,\,
\{A_K,\,{\vec n}_K\}}\, ,
\label{VINV}
\end{equation}
where $A_f, B_f$ represent both Lorentz and gauge indices of the various
fields.  Thanks to (\ref{VINV}), we notice that (see Fig.~4) if $K$
propagators are attached to a vertex, we do not have to sum over all their
$K$ independent twists, as one of them can be set to zero, simply giving
an extra factor of $N$. Notice that there is no need for the vertex $V$ to
be elementary, \textit{i.e.} to appear in the tree-level action.
This general result turns out to be useful in computing the Higgs 2-point
function. In this case it also holds for on-shell external lines, because $
\vec n =\vec 0$ is a fixed point of $Z_N$, and ${\mc P}$ acts as the identity
on the physical $\vec 0$ modes.

\begin{figure}[t]
\begin{center}
\begin{picture}(400,80)(0,5)
\put(10,0){\begin{picture}(150,50)(0,0)
\Vertex(75,25){3}
\ArrowLine(10,50)(75,25)
\BCirc(31,42){6}
\Text(32,42)[c]{\footnotesize $k_2$}
\ArrowLine(0,25)(75,25)
\BCirc(20,25){6}
\Text(21,25)[c]{\footnotesize $k_1$}
\ArrowLine(150,25)(75,25)
\BCirc(130,25){6}
\Text(131,25.5)[c]{\footnotesize $k_K$}
\ArrowLine(34,60)(75,25)
\BCirc(42,53){6}
\ArrowLine(150,40)(75,25)
\BCirc(138.5,38){6}
\DashCArc(75,25)(30,20,130){2}
\CBoxc(13,52)(17,7){White}{White}
\Text(4,19)[c]{\footnotesize ${\vec n}_1,\, a_1$}
\Text(8,54)[c]{\footnotesize ${\vec n}_2,\, a_2$}
\Text(152,19)[c]{\footnotesize ${\vec n}_K,\, a_K$}
\end{picture}}
\Text(200,30)[c]{$=\;N\;\times\;$}
\put(240,0){\begin{picture}(150,50)(0,0)
\Vertex(75,25){3}
\ArrowLine(10,50)(75,25)
\BCirc(31,42){6}
\Text(32,42)[c]{\footnotesize $k_2$}
\ArrowLine(0,25)(75,25)
\BCirc(20,25){6}
\Text(20,25)[c]{\footnotesize $0$}
\ArrowLine(150,25)(75,25)
\BCirc(130,25){6}
\Text(131,25.5)[c]{\footnotesize $k_K$}
\ArrowLine(34,60)(75,25)
\BCirc(42,53){6}
\ArrowLine(150,40)(75,25)
\BCirc(138.5,38){6}
\DashCArc(75,25)(30,20,130){2}
\CBoxc(13,52)(17,7){White}{White}
\Text(4,19)[c]{\footnotesize ${\vec n}_1,\, a_1$}
\Text(8,54)[c]{\footnotesize ${\vec n}_2,\, a_2$}
\Text(152,19)[c]{\footnotesize ${\vec n}_K,\, a_K$}
\end{picture}}
\end{picture}
\caption{\footnotesize Equivalence between interaction vertices
on a $\Z_N$ orbifold.}
\end{center}
\protect\label{f.2}
\end{figure}
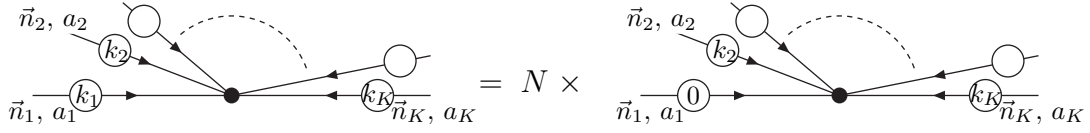

We extract the coefficients ${\mc C_k}$ by computing the 1-point
function of all the KK modes of $A_z^8$. We then work out also
the 2-point function for the zero-mode $h$ of the Higgs field,
defined as in (\ref{AzAzbar}), to extract its finite non-local
mass terms and to check the 1-point function computation. In order
to find an expression that can be directly compared with the computation
of 1- and 2-point functions for KK modes, we need to work out more
explicitly eq.~(\ref{O}). Using the mode expansion (\ref{FIexp}), we
easily find:
\begin{eqnarray}
\int \mathrm{d}z^2 {\mc L}_{\rm tad} = - \sum_{k=1}^{[N/2]} {\mc C}_k
\Bigg[\a\a\!\! \sum_{\vec n}\frac {1}{N_k} \sum_{i_k=1}^{N_k}
f_{\vec n}(z_{i_k}) \Big(p_{z,\vec n} A_{\bar z,\vec n}^{8}
- p_{\bar z,\vec n} A_{z,\vec n}^{8} \Big)\nn \\
\a\a\!\! +\, \frac{g_4}{\sqrt{V}}\Big(i f^{8+i-j}\Big)
h_ih_{j}^\dagger \Bigg]\, + \ldots \, ,
\label{OO}
\end{eqnarray}
where $p_{z,\vec m} = \frac {i}{\sqrt{2}} \lambda_{\vec m}$,
$p_{\bar z,\vec m} = - \frac {i}{\sqrt{2}} \bar\lambda_{\vec m}$
are the internal KK momenta, with $\lambda_{\vec m}$, $\bar\lambda_{\vec m}$
as in (\ref{lambda}), and the dots stand for all the remaining quadratic
couplings between all the KK excitations of $A_{z,+i}$ and $A_{\bar z,-j}$. Using
the identity
\begin{equation}
\frac {1}{N_k}\sum_{i_k=1}^{N_k}f_{\vec n}(z_{i_k})= \frac{1}{\sqrt{V}}
\delta_{(1-Z_N^k)^{-1}\vec n \in \ZZ^2}\,,
\end{equation}
valid for all the $T^2/\Z_N$ orbifolds, the contributions of the two
terms in (\ref{OO}) to the 1- and 2-point functions are found to be
\begin{eqnarray}
\a\a \langle A_{z,\vec n}^{8} \rangle = i p_{\bar z,\vec n}
\sum_{k=1}^{[N/2]} \frac {{\mc C}_k}{\sqrt{V}}\,
\delta_{(1-Z_N^k)^{-1}\vec n\in \ZZ^2}\,,
\label{1} \\
\a\a \langle h_i h_{j}^\dagger\rangle = g_4 f^{8+i-j}
\sum_{k=1}^{[N/2]} \frac {{\mc C}_k}{\sqrt{V}}\,.
\label{2}
\end{eqnarray}
Notice that the Higgs mass term arising from (\ref{2}) is sensible only to
the sum of the tadpole coefficients ${\mc C_k}$.

\subsection{1-point function}

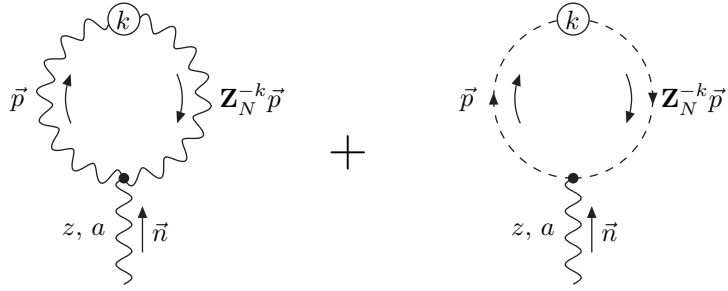
\begin{figure}[t]
\begin{center}
\begin{picture}(320,120)(0,10)
\put(0,0){\begin{picture}(150,120)(0,0)
\Vertex(75,50){2}
\PhotonArc(75,80)(30,-90,90){3}{9}
\PhotonArc(75,80)(30,90,270){3}{9}
\Photon(75,10)(75,50){-3}{4}
\BCirc(75,110){6}
\Text(75,110)[c]{\footnotesize $k$}
\LongArrow(82,22)(82,38)
\Text(89,30)[c]{\footnotesize $\vec n$}
\LongArrowArcn(75,80)(22,205,155)
\LongArrowArcn(75,80)(22,25,-25)
\Text(35,80)[c]{\footnotesize $\vec p$}
\Text(123,80)[c]{\footnotesize $\Z_{N}^{-k}\vec p$}
\Text(60,30)[c]{\footnotesize $z,\, a$}
\end{picture}}
\Text(160,60)[c]{\LARGE $+$}
\put(170,0){\begin{picture}(150,120)(0,0)
\Vertex(75,50){2}
\DashArrowArcn(75,80)(30,-90,90){3}
\DashArrowArcn(75,80)(30,90,270){3}
\Photon(75,10)(75,50){-3}{4}
\BCirc(75,110){6}
\Text(75,110)[c]{\footnotesize $k$}
\LongArrow(82,22)(82,38)
\Text(89,30)[c]{\footnotesize $\vec n$}
\LongArrowArcn(75,80)(22,205,155)
\LongArrowArcn(75,80)(22,25,-25)
\Text(35,80)[c]{\footnotesize $\vec p$}
\Text(120,80)[c]{\footnotesize $\Z_{N}^{-k}\vec p$}
\Text(60,30)[c]{\footnotesize $z,\, a$}
\end{picture}}
\end{picture}
\caption{\footnotesize The gauge and ghost contributions to the
1-point function $\langle A_{z,\vec n}^a \rangle$.}
\end{center}
\protect\label{Fey1p}
\end{figure}

According to the considerations made at the beginning of this section,
all the Feynman rules for the vertices are the standard ones, whereas the
propagator has to be replaced (see Fig.~5) by its twisted version, as
in Fig.~2. In the following we adopt the Feynman gauge, obtained through
the choice $\xi=1$ in the general gauge-fixing term
\begin{equation}
{\cal L}_{\rm gf} = -\frac{1}{2\xi} \sum_{a=1}^8
\Bigg[\partial_\mu A^{\mu,a} - \xi \big(\partial_z A_{\bar z}^a +
\partial_{\bar z} A_{z}^a\big) \Bigg]^2\,.
\label{xigauge}
\end{equation}

By $4D$ Lorentz invariance, a tadpole can be generated only for the
field components $A_z^a$ and $A_{\bar z}^a$. An explicit computation
of this tadpole shows that it has the form:
\begin{equation}
\langle A_{z,\vec n}^a \rangle = g_4 \sum_{k=0}^{N-1} \hat \xi_k^a
\sum_{\vec m} \int \frac{d^4p}{(2\pi)^4}
\frac{p_{\bar z,\vec m}}{p^2 - 2\,|p_{z,\vec m}|^2}
\delta_{ \vec n , (1-Z^{-k}_N)\vec m}\,,
\label{1general}
\end{equation}
where $\hat \xi_k^a$ are numerical coefficients depending on the kind
of field running in the loop and $p^2=p_\mu p^\mu$ is the $4D$ momentum
squared. The sector $k=0$ never contributes. For the sectors $k \neq 0$,
the $\delta$-function in KK space relates the internal momenta of the
virtual state to that of the external particle:
$p_{\bar z,\vec m} = (1-\tau^k)^{-1} p_{\bar z,\vec n}$.
We can then perform the sum over $m$, and we are left with the
condition $(1-Z^{-k}_N)^{-1} \vec n \in \ZZ^2$. Therefore, the
quadratically divergent part of eq.~(\ref{1general}) has the form of
eq.~(\ref{1}). This condition can easily be shown to be equivalent to
$(1-Z^{k-N}_N)^{-1} \vec n \in \ZZ^2$, so that in eq.~(\ref{1general})
the sector $N-k$ contributes just as the sector $k$. The two contributions
of these conjugate sectors can be paired, as expected, and simply yield
twice the real part of one of them (with the obvious exception of the sector
$k=N/2$ that, if present, must be counted only once). Finally, defining
the new coefficients $\xi_k^a= - 2^{1-\delta_{k,N/2}}\tau^{-k/2}N_k^{-1/2}
\hat\xi_k^a$, eq.~(\ref{1general}) can be rewritten in the more suggestive
form
\begin{equation}
\langle A_{z,\vec n}^a \rangle = - g_4 D(\Lambda) \sum_{k=1}^{[N/2]}
\,p_{\bar z,\vec n}\,\delta_{(1-Z^{k}_N)^{-1} \vec n \in \ZZ^2}\, {\rm Im}\,
\xi_k^ a\, + \dots \,,
\label{1calcolo}
\end{equation}
where
\begin{equation}
D(\Lambda) = i \int \!\! \frac{d^4p}{(2\pi)^4} \frac{1}{p^2}
= \frac{1}{16\pi^2} \Lambda^2\,.
\label{DLambda}
\end{equation}
The dots in (\ref{1calcolo}) stand for additional logarithmically divergent
and finite subleading corrections. These corrections are very similar
to those found in \cite{anomaly2} for the FI term in $5D$ SUSY theories
on $S^1/(\Z_2\times \Z_2^\prime)$, and are associated to interactions
involving additional internal derivatives $\partial_z \partial_{\bar z}$
acting on $F_{z\bar z}^8$. Notice also that in the presence of an additional
bulk mass term $M$ for the fields running in the loop, which is possible for
instance for scalar fields, eq.~(\ref{DLambda}) gets modified through the
simple substitution $\Lambda^2 \rightarrow \Lambda^2 - 2\, M^2 \ln
(\Lambda/M)$.

The contributions to $\xi_a^k$ of the gauge and ghost fields in the adjoint
representation, and of complex scalar or Weyl spinor fields in an arbitrary
representation $\R$ and with overall twist $g$, are found to be:
\begin{eqnarray}
(\xi_k^a)_{\rm gauge} \a
=\a \frac{-1}{NN_k^{1/2}}\bigg[5(\tau^{\frac k2}+\tau^{-\frac k2})-
(\tau^{\frac{3k}2}+\tau^{-\frac{3k}2})\bigg] \,
{\rm Tr}_{\rm adj} \Big[P^k\, t^{a} \Big] \,, \label{xi1} \\
(\xi_k^a)_{\rm scalar} \a=\a \frac{-2}{NN_k^{1/2}}\,g^k\,
(\tau^{\frac k2}+\tau^{-\frac k2})
{\rm Tr}_{\R} \Big[P^k\, t^{a} \Big] \,, \label{xi2} \\
(\xi_k^a)_{\rm fermion} \a=\a (4)\frac{2^{1-\delta_{k,N/2}}}{NN_k^{1/2}}\,
(g\tau^{\frac12})^k\,
{\rm Tr}_{\R} \Big[P^k\, t^{a} \Big] \,, \label{xi3}
\end{eqnarray}
where $(\xi_k^a)_{\rm gauge}$ also contains the ghost contribution. The gauge
trace appearing in the above coefficients is as expected to differ from zero
only for $a=8$, reflecting the fact that only a $U(1)$ tadpole is allowed by
the gauge symmetry. It is easily evaluated by recalling the definition of the
twist matrix $P_{\mc R}$, eq.~(\ref{Pierre}), and exploiting the decomposition
of the representation $\R$ under $SU(3) \rightarrow SU(2) \times U(1)$.
Denoting by $d_{\R_r}$ and $q_{\R_r}$ the dimensionality and the charge
under $Q_{\mc R}=\frac 1{\sqrt{3}}t_{\mc R}^8$ of the $r$-th component $\R_r$
in the decomposition $\R \rightarrow \oplus_r \R_r$, we find:
\begin{equation}
{\rm Tr}_{\mc R}\left[P^k t^8\right]= \sqrt{3} \sum_{\R_r}
d_{\R_r} q_{R_r} \tau^{2n_p(\frac{n_{\mc R}}{3}+q_{\R_r})k}\,.
\label{tracciagauge}
\end{equation}
Notice that the gauge contribution to the tadpole vanishes at 1-loop order
for the $\Z_2$ case. The same happens for any scalar or fermion contribution
in a real representation. This can be seen by using the relation (valid for
any $\Z_N$ orbifold):\footnote{Actually the scalar contribution in the $\Z_2$
model vanishes for any representation, not only for real ones.}
\begin{equation}
{\rm Tr}_{\R} \left[P^k t^8\right] =
- {\rm Tr}_{\R} \left[P^{-k} t^8\right]\,,\;\;
\mbox{if $\R$ real}.
\label{Real-P}
\end{equation}
This result is in agreement with that found in \cite{vonGersdorff2}, where
it was also generalized to the 2-loop case. On the contrary, for $N=3,4,6$,
there is always some tadpole coefficient that is non-vanishing for the
single-Higgs projections. The tadpole can only vanish for the zero-Higgs
cases $N=4$, $n_p=2$ and $N=6$, $n_p=3$, since they correspond to vanishing
Im ${\rm Tr}_{\rm adj} \Big[P^k\, t^{a} \Big]$.

The fermion contribution (\ref{xi3}) has a structure that resembles that of
the $4D$ mixed $U(1)$-gravitational anomaly induced at the fixed points by
$6D$ Weyl fermions. The structure of this anomaly can be understood using
Fujikawa's approach to anomalies, as was done in \cite{Scrucca:1999pp} for
string-derived orbifold models. The total contribution to localized mixed
$U(1)$-gravitational anomalies from a $6D$ fermion is proportional to
$\sum_{k=1}^{[N/2]} N_k\, {\rm Im}\,(\xi_k^ 8)_{\rm fermion}$. This expression
can be written as a projector over massless $4D$ fermions, weighted by their
$4D$ chirality, and is thus proportional to the sum over $U(1)$ charges of
the $4D$ chiral fermions. Notice, however, that $6D$ fermions always
contribute to the tadpole with the same sign, independently of their $6D$
chirality, as already noted for the $\Z_2$ case in \cite{vonGersdorff2}.
Since a flip of the $6D$ chirality amounts to a flip of the $4D$ one, this
implies that the sum over all possible fermion contributions to the tadpole
does not coincide with the total mixed $U(1)$-gravitational anomaly of the
$4D$ fermion spectrum, even when the factor $N_k$ can be factorized out of
the trace, as in the $\Z_2$ and $\Z_3$ models. This means that even when
the scalar and gauge contribution to the tadpole vanish, the requirements of
vanishing integrated tadpole and $U(1)$-gravitational anomaly cancellation
are in general independent constraints.

In order to relate the coefficients $\xi_k^8$ to the coefficients
${\mc C_k}$ appearing in (\ref{O}), we must compare eq.~(\ref{1calcolo})
with eq.~(\ref{1}), which have as expected the same structure.
The result is
\begin{equation}
{\mc C}_{k} = g_4\sqrt{V}D(\Lambda)\,{\rm Im}\,\xi_k^8\,.
\label{Ck}
\end{equation}
We summarize in Table~\ref{tab:tadpoles} the contribution of a Weyl fermion
to ${\mc C}_k$ for all possible twists and for the first few $SU(3)$
representations. Notice that the contribution of a fermion with twist
$g$ in the conjugate representation $\bar {\mc R}$ is equal to that of a
fermion twisted by $\bar g \bar \tau$ in the representation ${\mc R}$.
Similarly, a scalar in the $\bar {\mc R}$ with twist $g$ contributes as one
in the ${\mc R}$ with conjugate twist $\bar g$. The sum over all possible
twists for any scalar or fermion contribution always vanishes, since in this
case one reconstructs the matter content that would appear on the covering
torus, which cannot give rise to any localized divergence. We see that for
$N=3,4,6$ and for any choice of fermion representations, it is impossible to
cancel the total (gauge+ghost+fermion) one-loop contribution to each tadpole
coefficient\footnote{Our result for the gauge+ghost one-loop contribution
to the tadpole in the $\Z_4$ model is in disagreement with the result of
\cite{Csaki:2002ur}, where it was found to vanish.}, although one can obtain
their global cancellation, namely the cancellation of their integral over
the compact space $\sum_k {\mc C}_k = 0$. This seems to be possible, without
scalars, only for $\Z_4$ with an odd number of $6D$ Weyl fermions in suitable
representations. If one includes scalars, an accidental local one-loop
cancellation of the tadpole is possible, but in this case one needs a symmetry
to protect the mass $M$ of the $6D$ scalars, which is otherwise expected to
be at the cut-off scale $\Lambda$, and reintroduce a quadratic sensitivity to
the latter.

\begin{table}[h]
\begin{tabular}{ll}
${}$ \hspace{-9pt}
\begin{tabular}{|c|c|c|c|c|c|c|}
\cline{1-3}
$\Z_2$ \b & $1$ & $\tau$ & \multicolumn{4}{|c}{\ }\\
\cline{1-3}
$c_1({\bf 3})$ \d
&$-4 $
&$4 $
&\multicolumn{4}{|c}{\ }
\\
$c_1({\bf 6})$ \d
&$4 $
&$-4 $
&\multicolumn{4}{|c}{\ }
\\
$c_1({\bf 8})$ \d
&$0 $
&$0 $
&\multicolumn{4}{|c}{\ }
\\
$\!c_1({\bf 10})\!$ \d
&$-12 $
&$12 $
&\multicolumn{4}{|c}{\ }
\\
\cline{1-3}
\end{tabular}
&
\hspace{-20pt}
\begin{tabular}{|c|c|c|c|c|c|c|}
\cline{1-4}
$\Z_3$ \b & $1$ & $\tau$ & $\tau^2$ & \multicolumn{3}{|c}{\ }\\
\cline{1-4}
$c_1({\bf 3})$ \d
&$-4$
&$-4$
&$8$
&\multicolumn{3}{|c}{\ }
\\
$c_1({\bf 6})$ \d
&$-20$
&$16$
&$4$
&\multicolumn{3}{|c}{\ }
\\
$c_1({\bf 8})$ \d
&$12$
&$-24$
&$12$
&\multicolumn{3}{|c}{\ }
\\
$\!c_1({\bf 10})\!$ \d
&$12$
&$12$
&$-24$
&\multicolumn{3}{|c}{\ }
\\
\cline{1-4}
\end{tabular}
\vspace{13pt} \nn \\
${}$ \hspace{-13pt}
\raisebox{34pt}{
\begin{tabular}{|c|c|c|c|c|c|c|}
\cline{1-5}
$\Z_4$ \b & $1$ & $\tau$ & $\tau^2$ & $\tau^3$ & \multicolumn{2}{|c}{\ }
\\
\cline{1-5}
$c_1({\bf 3})$ \d
&$0$
&$-8$
&$0$
&$8$&\multicolumn{2}{|c}{\ }\\
$c_1({\bf 6})$ \d
&$-24$
&$-16$
&$24$
&$16$&\multicolumn{2}{|c}{\ }\\
$c_1({\bf 8})$ \d
&$24$
&$-24$
&$-24$
&$24$&\multicolumn{2}{|c}{\ }\\
$\!c_1({\bf 10})\!$ \d
&$-48$
&$24$
&$48$
&$-24$&\multicolumn{2}{|c}{\ }\\
\cline{1-5}
$c_2({\bf 3})$ \d
&$-4$
&$4$
&$-4$
&$4$&\multicolumn{2}{|c}{\ }
\\
$c_2({\bf 6})$ \d
&$4$
&$-4$
&$4$
&$-4$&\multicolumn{2}{|c}{\ }
\\
$c_2({\bf 8})$ \d
&$0$
&$0$
&$0$
&$0$&\multicolumn{2}{|c}{\ }
\\
$\!c_2({\bf 10})\!$ \d
&$-12$
&$12$
&$-12$
&$12$&\multicolumn{2}{|c}{\ }
\\
\cline{1-5}
\end{tabular}
}
&
\hspace{-20pt}
\begin{tabular}{|c|c|c|c|c|c|c|}
\cline{1-7}
$\Z_6$ \b & $1$ & $\tau$ & $\tau^2$ & $\tau^3$ & $\tau^4$ & $\tau^5$ \\
\cline{1-7}
$c_1({\bf 3})$ \d
&$4$
&$-4$
&$-8$
&$-4$
&$4$
&$8$\\
$c_1({\bf 6})$ \d
&$-4$
&$-32$
&$-28$
&$4$
&$32$
&$28$\\
$c_1({\bf 8})$ \d
&$36$
&$0$
&$-36$
&$-36$
&$0$
&$36$\\
$\!c_1({\bf 10})\!$ \d
&$-60$
&$-84$
&$-24$
&$60$
&$84$
&$24$\\
\cline{1-7}
$c_2({\bf 3})$ \d
&$-4$
&$-4$
&$8$
&$-4$
&$-4$
&$8$\\
$c_2({\bf 6})$ \d
&$-20$
&$16$
&$4$
&$-20$
&$16$
&$4$\\
$c_2({\bf 8})$ \d
&$12$
&$-24$
&$12$
&$12$
&$-24$
&$12$\\
$\!c_2({\bf 10})\!$ \d
&$12$
&$12$
&$-24$
&$12$
&$12$
&$-24$\\
\cline{1-7}
$c_3({\bf 3})$ \d
&$-4$
&$4$
&$-4$
&$4$
&$-4$
&$4$\\
$c_3({\bf 6})$ \d
&$4$
&$-4$
&$4$
&$-4$
&$4$
&$-4$\\
$c_3({\bf 8})$ \d
&$0$
&$0$
&$0$
&$0$
&$0$
&$0$\\
$\!c_3({\bf 10})\!$ \d
&$-12$
&$12$
&$-12$
&$12$
&$-12$
&$12$\\
\cline{1-7}
\end{tabular}
\end{tabular}
\caption{\footnotesize The contribution to the tadpole coefficients
${\mc C}_k$ from Weyl fermions for various representations and all
choices of the phase $g$. We report the
quantity $c_k=\sqrt{3}\,N\,{\rm Im}[\xi_k^8]$, which for the gauge
contribution is given by $c_1=0$ for $\Z_2$, $c_1=-21$ for $\Z_3$,
$c_1=-36$ and $c_2=0$ for $\Z_4$, and $c_1=-45$, $c_2=-21$ and
$c_3=0$ for $\Z_6$. In all cases, we are considering the projection with
$n_p=1$, giving single-Higgs models for $N\neq 2$.}
\label{tab:tadpoles}
\end{table}

\subsection{2-point function}

\begin{figure}[t]
\begin{center}
\begin{picture}(410,80)(0,0)
\put(0,0){\begin{picture}(120,80)(0,0)
\Vertex(35,40){2}
\Vertex(85,40){2}
\PhotonArc(60,40)(25,0,90){3}{4}
\PhotonArc(60,40)(25,90,180){3}{4}
\PhotonArc(60,40)(25,180,360){3}{8}
\Photon(5,40)(35,40){-3}{3.5}
\Photon(85,40)(115,40){-3}{3.5}
\BCirc(60,65){6}
\Text(60,65)[c]{\footnotesize $k$}
\LongArrow(8,35)(18,35)
\LongArrow(112,35)(102,35)
\Text(11,27)[c]{\footnotesize $\vec 0$}
\Text(109,27)[c]{\footnotesize $\vec 0$}
\LongArrowArcn(60,40)(20,155,115)
\LongArrowArcn(60,40)(20,65,25)
\LongArrowArcn(60,40)(20,-60,-120)
\Text(10,47)[c]{\footnotesize $z,\, +i$}
\Text(110,47)[c]{\footnotesize ${\bar z},\, -j$}
\Text(39,65)[c]{\footnotesize $\vec p$}
\Text(85,65)[c]{\footnotesize $\Z_{N}^{-k}\vec p$}
\Text(60,6)[c]{\footnotesize $\Z_{N}^{-k}\vec p$}
\end{picture}}
\Text(132.5,40)[c]{$+$}
\put(145,0){\begin{picture}(120,80)(0,0)
\Vertex(35,40){2}
\Vertex(85,40){2}
\DashArrowArcn(60,40)(25,90,0){3}
\DashArrowArcn(60,40)(25,180,90){3}
\DashArrowArcn(60,40)(25,360,180){3}
\Photon(5,40)(35,40){-3}{3.5}
\Photon(85,40)(115,40){-3}{3.5}
\BCirc(60,65){6}
\Text(60,65)[c]{\footnotesize $k$}
\LongArrow(8,35)(18,35)
\LongArrow(112,35)(102,35)
\Text(11,27)[c]{\footnotesize $\vec 0$}
\Text(109,27)[c]{\footnotesize $\vec 0$}
\LongArrowArcn(60,40)(20,155,115)
\LongArrowArcn(60,40)(20,65,25)
\LongArrowArcn(60,40)(20,-60,-120)
\Text(10,47)[c]{\footnotesize $z,\, +i$}
\Text(110,47)[c]{\footnotesize ${\bar z},\, -j$}
\Text(39,65)[c]{\footnotesize $\vec p$}
\Text(85,65)[c]{\footnotesize $\Z_{N}^{-k}\vec p$}
\Text(60,6)[c]{\footnotesize $\Z_{N}^{-k}\vec p$}
\end{picture}}
\Text(277.5,40)[c]{$+$}
\put(290,0){\begin{picture}(120,80)(0,0)
\Vertex(60,20){2}
\PhotonArc(60,45)(25,-90,90){3}{8}
\PhotonArc(60,45)(25,90,270){3}{8}
\Photon(5,20)(60,20){-3}{5}
\Photon(60,20)(115,20){-3}{5}
\BCirc(60,70){6}
\Text(60,70)[c]{\footnotesize $k$}
\LongArrow(8,14)(18,14)
\LongArrow(112,14)(102,14)
\Text(11,6)[c]{\footnotesize $\vec 0$}
\Text(109,6)[c]{\footnotesize $\vec 0$}
\LongArrowArcn(60,45)(20,210,150)
\LongArrowArcn(60,45)(20,30,-30)
\Text(10,27)[c]{\footnotesize $z,\, +i$}
\Text(110,27)[c]{\footnotesize ${\bar z},\, -j$}
\Text(30,45)[c]{\footnotesize $\vec p$}
\Text(100,45)[c]{\footnotesize $\Z_{N}^{-k}\vec p$}
\end{picture}}
\end{picture}
\caption{\footnotesize  The gauge and ghost contributions to the 2-point
function $\langle h_{i}h_{j}^\dagger \rangle$.}
\end{center}
\protect\label{Fey2p}
\end{figure}
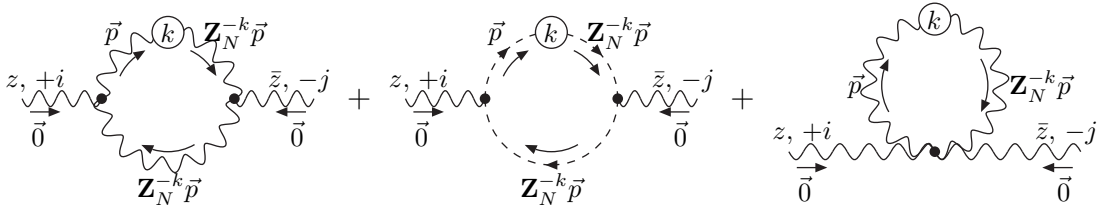

We now compute the one-loop 2-point function for the Higgs field, at
zero external $4D$ and KK momentum. Contrarily to the 1-point function,
which we have computed for any external KK momentum, this correlation
gives us information only on the form of the operator (\ref{O})
integrated over the compact space (see eq.~(\ref{2})). Nevertheless, it
provides an important independent check of the 1-point function computation
and also allows the extraction of the finite non-local contributions to the
Higgs mass.

Thanks to the property displayed in Fig.~4,
each of the diagrams contributing to the one-loop Higgs mass contains only
one twisted propagator with twist $k$. The diagrams with $k=0$ give a finite
contribution, which reproduces up to a $1/N$ factor the result that would be
obtained for a theory on the covering torus $T^2$. The remaining contributions
arising from the diagrams with the insertion of a propagator with $k\neq 0$
are instead divergent. Owing to momentum conservation at the vertices, the
internal KK momentum in the twisted internal lines has to vanish (see Fig.~6).
The general structure of the Higgs 2-point function is then given
by:\footnote{Equation (\ref{2calcolo}) is valid also for the $\Z_2$ model,
where two Higgs fields are present. In this case, there are additional
2-point correlators that we neglect. See {\em e.g.} \cite{Antoniadis:2001cv}.}
\begin{equation}
\langle h_i\, h_{j}^\dagger\rangle =
g_4^2 f_{8}^{\; +i-j} \xi_{\rm div}^8\, D(\Lambda)
+ i g_4^2\,\delta_{ij}\,\xi_{\rm fin}^8\, F(R) \,,
\label{2calcolo}
\end{equation}
with $D(\Lambda)$ as in (\ref{DLambda}) and
\begin{equation}
F(R) = i\int \frac {d^4p_4}{(2\pi)^4} \sum_{\vec n\in \ZZ^2}
{\ds \frac{p^2}{(p^2 - 2 |p_{z,\vec n}|^2)^2}} =
\frac {U_2}{4 \pi^5 R^2} \sum_{\vec n \neq \vec 0}
|n_1 + U n_2|^{-4}\,.
\label{F-finito}
\end{equation}

It is straightforward to compute the diagrams controlling the divergent part.
Note that ghosts do not contribute, because their coupling to the Higgs
is proportional to the KK momentum. Thanks to the identities
\begin{eqnarray}
{\rm Tr}_\R \left[t^{+i}t^{-j}P^k\right]=\a\a
\tau^l{\rm Tr}_\R \left[t^{-j}t^{+i}P^k\right]\, , \nn \\
{\rm Tr}_\R \left[t^{-j}t^{+i}P^k\right]=\a\a -if_{8}^{\;+i-j}
\frac 1{1-\tau^l} {\rm Tr}_\R\left[t^{8} P^k\right]\,,
\label{Gauge-Id1}
\end{eqnarray}
the result can be rewritten as
\begin{equation}
\xi_{\rm div}^8 = \sum_{k=1}^{[N/2]} {\rm Im}\, \xi_k^8\,,
\label{rell}
\end{equation}
where the $\xi_k^8$'s turn out to precisely match the expressions
(\ref{xi1})--(\ref{xi3}) extracted from the 1-point function computation.
This result represents a non-trivial check of that computation. Indeed,
comparing eq.~(\ref{2calcolo}) with eq.~(\ref{2}), we deduce that
\begin{equation}
\sum_{k=1}^{[N/2]} {\mc C}_{k} =  g_4 \sqrt{V} D(\Lambda)\,\xi_{\rm div}^8\,
\end{equation}
which is compatible with the result in eq.~(\ref{Ck}) thanks to the
relation (\ref{rell}).

The diagrams contributing to the finite part can be computed as well,
and the coefficients of the finite part are found to be given by:
\begin{eqnarray}
(\xi_{\rm fin}^8)_{\rm gauge} \a=\a 2 \frac 4{N} C({\rm Adj})\,, \\
(\xi_{\rm fin}^8)_{\rm scalar} \a=\a \frac 4{N} C(\R)\,, \\
(\xi_{\rm fin}^8)_{\rm fermion} \a=\a - 2 \frac 4{N} C(\R)\,,
\end{eqnarray}
in terms of the quadratic Casimir $C(\R)$ of the representation
$\R$, defined by the relation
${\rm Tr}_{\R}\left[t^a t^b\right]=C(\R)\delta^{ab}$,
so that $C({\rm Fund}) = \frac 12$ and $C({\rm Adj}) = 3$.

\section{Phenomenological implications}

In sections 3 and 4 we have shown how $6D$ gauge theories on orbifold
models can lead to a beautiful prediction for the Higgs mass, but at
the same time they are affected by a quadratic divergence arising
from a localized tadpole term. It is thus natural to try to understand
whether and to what extent such models can be considered for realistic
model building.

One of the main generic problems in models with gauge--Higgs unification
is how to accommodate the standard matter fields. A possibility is to
introduce them at the fixed points of the orbifold. Standard Yukawa
couplings cannot be directly introduced, because they would violate the
higher-dimensional symmetry. However, effective non-local Yukawa couplings
can be generated by introducing mixings between the matter fields and
additional heavy fermions in the bulk, which are then integrated out
\cite{Csaki:2002ur,Scrucca:2003ra}. In this situation, the localized matter
field will in general influence the one-loop tadpole (\ref{tadpolo}),
but since this requires mixing insertions, only logarithmic divergences
can be induced. The weak mixing angle, whose value in the basic $SU(3)$
model is too large, can be fixed by introducing additional $U(1)$ gauge
fields in the bulk (see {\em e.g.} \cite{Antoniadis:2001cv,Scrucca:2003ra}),
as mentioned in section 2.

The real issue is that the presence of the quadratically divergent
term (\ref{O}) can destabilize the electroweak scale. It must therefore
be understood how much (if any) progress has been achieved with
respect to the SM, as far as the little hierarchy problem is
concerned. The abelian and non-abelian components of the localized
operator (\ref{O}) induce respectively a non-trivial background for the
field $A_z^8$ and a mass term for the Higgs doublet $A_z^{-i}$. The latter
can generate not only a mass term for the $4D$ Higgs field, but also
mixings between all its KK partners.  These mixings can be neglected only if
their magnitude is much smaller than $1/R$, the typical mass of KK
modes. In our case, $\mc C_k > 1/R$ (see below) and the effect of
all these mixings, as well as that of the non-trivial background for
$A_z^8$, must be taken into account. In order to see if and how much the
EWSB scale is sensitive to this divergence, one has to compute
the background value of $A_z^8$ and study the quantum fluctuations
around it, to get the physical masses of the various fields, in
particular for $A_z^{-i}$. Luckily, a similar analysis has already been
performed in \cite{Lee:2003mc}, where the effect of localized FI terms
in 6D orbifold models has been studied (see also
\cite{Ghilencea:2001bw,GrootNibbelink:2003gb}). As already mentioned,
the tadpole (\ref{O}) can be interpreted as a FI term in SUSY theories;
this suggests a correspondence that allows a study of its physical
consequences even in our non-SUSY set-up. The background induced by the
tadpole can be explicitly found as follows. If one sets to zero all
$4D$ gauge fields, the effective potential one obtains for the scalar
fields $A_z^a$, in the unitary gauge $\xi\rightarrow \infty$ in
(\ref{xigauge}), can be written, up to some irrelevant constant terms, as
\begin{equation}
V = \frac 12 \sum_{a=1}^3|F_{z
\bar z}^a|^2 + |F_{z \bar z}^{-i}|^2+ \frac 12 \Big|F_{z \bar z}^8 -
i\sum_{k=1}^{[N/2]}\frac{{\mc C_k}}{N_k} \sum_{i_k=1}^{N_k}
\delta^{(2)}(z-z_{i_k})\Big|^2 \,.
\label{potTad}
\end{equation}
The potential (\ref{potTad}) is a sum of squares and thus, as happens for
the $D$-term potential of SUSY theories, configurations where it vanishes
are automatically consistent classical backgrounds. In the particular case
where the tadpole globally vanishes, that is $\sum_k {\mc C_k} =0$, the
background value of the fields can be determined by proceeding as in
\cite{Lee:2003mc}. The result is that $\langle A_z^8\rangle = i \partial_z
W$, where the function $W$ is a linear combination of scalar Green functions
on the internal $T^2$. The existence of a zero-mode solution $A_{z,0}^{-i}$
for the field $A_z^{-i}$ in the presence of this background is ensured by the
existence of a solution to the first-order equation
\begin{equation}
F_{\bar z z}^{-i}= \bigg(\partial_{\bar
z}+ i g_6 \tan \theta_W \frac 12 \langle A_z^{8}\rangle \bigg)
A_{z,0}^{-i} = 0\,.
\label{F=0}
\end{equation}
We refer the reader to \cite{Lee:2003mc} for a detailed analysis of the
profile of the zero mode wave function in the internal space. The above
reasoning shows that a globally vanishing tadpole does not give rise to
any quadratic divergence in the Higgs mass parameter $\mu^2$ appearing in
(\ref{Vheff}). In other words, a globally vanishing tadpole is harmless
for EWSB, which is governed by the finite non-local contributions to the
2-point function, proportional to (\ref{F-finito}), which for simplicity
we have neglected in these simple lines of arguments. In the presence of
globally vanishing one-loop tadpole, the little hierarchy problem is then
solved. The non-trivial profile for the Higgs field,
however, induces large corrections to (\ref{pred}), since this ratio
depends on the integral of the Higgs profile in the internal directions.
As mentioned in section 3, such corrections are estimated to be
of $O(1)$ and thus might significantly alter the tree-level result
(\ref{pred}).

To be more quantitative, we can rely on the higher-dimensional
generalization of the NDA \cite{NDA}.
We denote in short by $l_4 = 16 \pi^2$ and $l_6=128\pi^3$ the $4D$ and $6D$
loop factors. The relation between the cut-off scale $\Lambda$ and
the compactification scale $1/R$ is then estimated to be
$\Lambda \sim g_4^{-1} (2 \pi R)^{-1} \sqrt{l_6}$, which for the EW
coupling yields $\Lambda \sim 10/R$. In this way we obtain an estimate
for the tadpole coefficient ${\mc C_k}$ that is in agreement with
the direct one-loop result reported in (\ref{Ck}) and of order
$l_6/(2\pi R g_4 l_4)> 1/R$, as mentioned. On the other hand, the value
of $\mu^2$ in eq.~(\ref{Vheff}) induced by finite non-local corrections
is of order $\mu^2 \sim g_4^2/(l_4 R^{2})$, and from (\ref{mh-mw}) one
estimates $1/R \sim 1$ TeV and $\Lambda \sim 10$ TeV, which are compatible
with present experimental bounds in a natural way. On the other hand, for
a globally non-vanishing tadpole, it is reasonable to expect to have
effectively $\mu^2 \sim g_4^2 \Lambda^2/l_4$. From (\ref{mh-mw}) one
now estimates $\Lambda \sim 1$ TeV, corresponding to $1/R \sim$ 100
GeV. The amount of fine-tuning that is needed in this case is about
the same as in the SM, and there is no progress concerning the little
hierarchy problem.

Summarizing, we have shown that there exists a class of $6D$ $\Z_N$
orbifold models with gauge--Higgs unification that lead to a single Higgs
doublet with the tree-level prediction $m_H = 2 m_W$, and for which the
EWSB scale is sufficiently stable, provided that the one-loop integrated
tadpole vanishes. Although we have not provided in this paper a complete
and realistic $6D$ model with gauge--Higgs unification, which would require
in particular to find an anomaly-free fermion spectrum with a globally
vanishing tadpole, we think that it will be very interesting to analyse the
phenomenological aspects of the single Higgs $T^2/\Z_N$ models discussed
in this paper.

\acknowledgments

We would like to thank
S.~Bertolini, R.~Contino, A.~Pomarol, M.~Quiros, R.~Rattazzi, A.~Romanino and
A.~Schwimmer for many fruitful discussions. This research work was partly
supported by the European Community through a Marie Curie fellowship and
the RTN network contracts HPRN-CT-2000-00131 and HPRN-CT-2000-00148.
M.S., L.S. and A.W. acknowledge CERN, and M.S. and A.W. acknowledge
Rome University ``La Sapienza'', where part of this work was done.

\end{document}